\documentclass[12pt]{article}
\newtheorem{stat}{Statement}[section]
\newcommand{\bstat}{\begin{stat}}
\newcommand{\estat}{\end{stat}}
\makeatletter
\def\section{\@startsection {section}{1}{\z@}{-3.5ex plus -1ex minus
 -.2ex}{2.3ex plus .2ex}{\large\bf}}
\def\subsection{\@startsection{subsection}{2}{\z@}{-3.25ex plus -1ex 
minus -.2ex}{1.5ex plus .2ex}{\normalsize\bf}}
\makeatother
\makeatletter

\@addtoreset{equation}{section}

\makeatother

\newsavebox{\uuunit}
\sbox{\uuunit}
    {\setlength{\unitlength}{0.825em}
     \begin{picture}(0.6,0.7)
        \thinlines
        \put(0,0){\line(1,0){0.5}}
        \put(0.15,0){\line(0,1){0.7}}
        \put(0.35,0){\line(0,1){0.8}}
       \multiput(0.3,0.8)(-0.04,-0.02){12}{\rule{0.5pt}{0.5pt}}
     \end {picture}}

\def\IP{\relax{\rm I\kern-.18em P}}

\begin{document}

\font\cmss=cmss10 \font\cmsss=cmss10 at 7pt
\def\twomat#1#2#3#4{\left(\matrix{#1 & #2 \cr #3 & #4}\right)}
\def\inbar{\vrule height1.5ex width.4pt depth0pt}
\def\IC{\relax\,\hbox{$\inbar\kern-.3em{\rm C}$}}
\def\IG{\relax\,\hbox{$\inbar\kern-.3em{\rm G}$}}
\def\IB{\relax{\rm I\kern-.18em B}}
\def\ID{\relax{\rm I\kern-.18em D}}
\def\IL{\relax{\rm I\kern-.18em L}}
\def\IF{\relax{\rm I\kern-.18em F}}
\def\IH{\relax{\rm I\kern-.18em H}}
\def\II{\relax{\rm I\kern-.17em I}}
\def\IN{\relax{\rm I\kern-.18em N}}
\def\IP{\relax{\rm I\kern-.18em P}}
\def\IQ{\relax\,\hbox{$\inbar\kern-.3em{\rm Q}$}}
\def\bfzero{\relax\,\hbox{$\inbar\kern-.3em{\rm 0}$}}
\def\IK{\relax{\rm I\kern-.18em K}}
\def\IG{\relax\,\hbox{$\inbar\kern-.3em{\rm G}$}}
 \font\cmss=cmss10 \font\cmsss=cmss10 at 7pt
\def\IR{\relax{\rm I\kern-.18em R}}
\def\ZZ{\relax\ifmmode\mathchoice
{\hbox{\cmss Z\kern-.4em Z}}{\hbox{\cmss Z\kern-.4em Z}}
{\lower.9pt\hbox{\cmsss Z\kern-.4em Z}}
{\lower1.2pt\hbox{\cmsss Z\kern-.4em Z}}\else{\cmss Z\kern-.4em
Z}\fi}
\def\bfone{\relax{\rm 1\kern-.35em 1}}
\def\dop{{\rm d}\hskip -1pt}
\def\real{{\rm Re}\hskip 1pt}
\def\trace{{\rm Tr}\hskip 1pt}
\def\ii{{\rm i}}
\def\diag{{\rm diag}}
\def\sch#1#2{\{#1;#2\}}
\def\bfone{\relax{\rm 1\kern-.35em 1}}
\font\cmss=cmss10 \font\cmsss=cmss10 at 7pt
\def\a{\alpha} \def\b{\beta} \def\d{\delta}
\def\e{\epsilon} \def\c{\gamma}
\def\G{\Gamma} \def\l{\lambda}
\def\L{\Lambda} \def\s{\sigma}
\def\cA{{\cal A}} \def\cB{{\cal B}}
\def\cC{{\cal C}} \def\cD{{\cal D}}
\def\cF{{\cal F}} \def\cG{{\cal G}}
\def\cH{{\cal H}} \def\cI{{\cal I}}
\def\cJ{{\cal J}} \def\cK{{\cal K}}
\def\cL{{\cal L}} \def\cM{{\cal M}}
\def\cN{{\cal N}} \def\cO{{\cal O}}
\def\cP{{\cal P}} \def\cQ{{\cal Q}}
\def\cR{{\cal R}} \def\cV{{\cal V}}
\def\cW{{\cal W}}
\newcommand{\be}{\begin{equation}}
\newcommand{\ee}{\end{equation}}
\newcommand{\bea}{\begin{eqnarray}}
\newcommand{\eea}{\end{eqnarray}}
\let\la=\label \let\ci=\cite \let\re=\ref
%
%
%
\def\crr{\crcr\noalign{\vskip {8.3333pt}}}
\def\tilde{\widetilde}
\def\bar{\overline}
\def\us#1{\underline{#1}}
\def\IE{\relax{{\rm I\kern-.18em E}}}
\def\cE{{\cal E}}
\def\rt{{\cR^{(3)}}}
\def\IGam{\relax{{\rm I}\kern-.18em \Gamma}}
\def\IGa{\IA}
\def\ii{{\rm i}}
\def\beq{\begin{equation}}
\def\eeq{\end{equation}}
\def\beqa{\begin{eqnarray}}
\def\eeqa{\end{eqnarray}}
\def\nn{\nonumber}

\begin{titlepage}
\setcounter{page}{0}

\begin{flushright}

SISSA REF 129/99/EP

SWAT/243

\end{flushright}

\vskip 30pt

\begin{center}

{\Large \bf Regular R--R and NS--NS BPS black holes}

\vskip 20pt

{\large Matteo Bertolini$^{a,}$\footnote{Address after November 1st: {\it NORDITA, 
Blegdamsvej 17, DK-2100 Copenhagen, Denmark}.\\ Supported in part by EEC under TMR 
contracts ERBFMRX--CT96--0045 and ERBFMRX--CT96--0012. } and Mario Trigiante$^{b}$}

\vskip 15pt

{\small
{\it $^a$International School for Advanced Studies ISAS-SISSA and INFN \\
Sezione di Trieste, Via Beirut 2-4, 34013 Trieste, Italy} \\

\vskip 5pt
{\it $^b$Department of Physics, University of Wales Swansea, Singleton Park \\
Swansea SA2 8PP, United Kingdom}
}
\end{center}

\vskip 30pt

\begin{abstract}
We show in a precise group theoretical fashion how the generating
solution  of regular BPS black holes of $N=8$ supergravity, which is
known to be a  solution also of a simpler $N=2$ $STU$ model
truncation,  can be characterized as purely NS--NS or R--R charged
according to the way the corresponding $STU$ model is embedded in the
original $N=8$ theory. Of particular interest is the  class of
embeddings which yield regular BPS black hole solutions carrying only
R--R charge and whose microscopic description can possibly  be given
in terms of bound states of  D--branes only. The microscopic
interpretation of the bosonic fields in this class of $STU$ models
relies on the solvable Lie algebra (SLA) method. In the present
article we improve this mathematical technique in order to provide two
distinct descriptions for type IIA and type IIB theories and an
algebraic characterization of $S\times T$--dual embeddings within the
$N=8,\,d=4$ theory. This analysis will be applied to the particular
example of a four parameter (dilatonic) solution of which both the 
full macroscopic and microscopic descriptions will be worked out.
\end{abstract}

\vskip 80pt

\begin{flushleft}
{\footnotesize
e-mail: teobert@sissa.it, m.trigiante@swansea.ac.uk}
\end{flushleft}

\end{titlepage}

\section{Introduction}

After the characterization of $D$--branes as R--R charged non--perturbative states of 
closed superstring theory \cite{pol}, there have been successful
microscopic computations of the entropy of some extremal and
non--extremal black hole configurations which reproduced, at the
microscopic level, the expected Beckenstein--Hawking behavior \cite{strvaf}-\cite{msw}.
However, despite these encouraging results, an open problem, nowadays, is still to
find a general recipe to give this correspondence based on first
principles other than specific computations. Actually, while gravity seems
to describe the quantum properties of all black holes in a unified but
incomplete way (since it provides only a macroscopic description), string theory seems to give 
nice answers but losing the
unified character of the properties of different black holes. 
It would be necessary to find a microscopic but still
unified way of describing black hole physics in the context of string
theory. In the last two years there have been in fact various attempts,
especially within the $AdS/CFT$ correspondence \cite{maldaac}, to give an
answer to this question relying on some general principles but a definite
answer has not been found yet. For recent progress in this direction see
for example \cite{str1}-\cite{finn1}.

A complementary strategy, which could be helpful in this respect, is to
take advantage of the $U$--duality properties of BPS black holes (like for
instance the invariance of the entropy under $U$--duality transformations)
and use them to infer the common underlined structure of very different
black holes sharing the same entropy. In particular, if one is able to
give a precise correspondence between at least one macroscopic solution
(ultimately the 5 parameters generating one) and its microscopic
description, and is {\it really} able to act on it via duality
transformations, one could derive the
microscopic stringy description of any macroscopic solution. Even of
those solutions (as the pure NS--NS ones) for which a microscopic
entropy counting has not been achieved yet\footnote{See \cite{wl1} for earlier 
qualitative results on the microscopic interpretation of NS--NS black hole entropy.}. The
possibility of having a control, both at macroscopic and microscopic
level, on all regular black holes with a given entropy could
shed further light on the very conceptual basis of the microscopic entropy
within string theory. This is the spirit this
paper relies on. 

Some time ago it has been shown that the generating solution of
regular $N=8$ black holes can be characterized as a solution within
the $STU$ model \cite{mp1}. Such a model is a $N=2$ truncation of the
$N=8$ original theory \cite{cj}. Any regular BPS black hole solution
within this  latter model preserving 1/8 of the original
supersymmetries (as any other regular BPS black hole solution should
be: indeed the 1/2 and 1/4  solutions of $N=8, \,D=4$ supergravity
have vanishing entropy, \cite{mp2}) is, modulo U--duality
transformations, a 1/2 BPS soliton of the $N=2$ theory.  This
important result enables  one to concentrate on the simpler
structure of the $STU$ model, and then to generate more general  (and
complicated) solutions by $U$--duality transformations. Without
specification of the proper embedding in the mother $N=8$ theory these
$STU$ model solutions can be NS--NS, R--R or of a mixed nature. This
distinction, from the $4$ dimensional point of view, relies on the
identification of the relevant (dimensionally reduced) $10$
dimensional fields which enter dynamically in the solution.  Of
particular interest will be the solutions which carry only  charge
with respect to vector potentials deriving from the dimensional
reduction of 10 dimensional R--R forms. In principle these BPS black
holes can be described microscopically by means of a suitable system
of D--branes compactified on an internal manifold (in the weak
coupling limit). The choice of a particular point in the moduli space
of the theory defining the values of the scalar fields of the solution
at radial infinity, although uninfluential as far as the macroscopic
properties of the black hole are concerned ({\it no--hair} theorem),
determines the particular D--brane configuration corresponding to the given 
solution in the opposite regime of the string coupling constant. We
shall consider in what  follows two main embeddings of the $STU$ model
in the $N=8$ theory: one in which the vector fields have a
10--dimensional interpretation in terms of NS--NS fields as opposed to
the other in which the vector fields have a R--R origin. These two
categories of embeddings are {\it equivalence classes} with respect to
the action of $S\times T$ duality (to be defined more rigorously in
the sequel and in particular in the appendix). Within the equivalence
class of the R--R charged solutions, two particular representative
embeddings will be considered: one in which the six real scalars of
the $STU$ model (three dilatons and three axions) come only from the
compactified components of the metric ($G_{ij}$) and an other in which
the three axions derive from compactified components of the
anti--symmetric tensor in 10 dimensions ($B_{ij}$). As we shall see,
the two embeddings are related by $T$--duality along three orthogonal
directions  of the internal torus. The former will be described in the
framework of type IIB theory and the latter in the framework of type
IIA theory. Black hole solutions in the above type IIB embedding could
be possibly interpreted microscopically in terms of a system of D3--branes 
at angles, while those in the type IIA embedding may  be
interpreted in terms of D0 and D4--branes with magnetic fluxes on the
world volume of the latter (since the fields $B_{ij}$ are coupled to
the D4--branes in a gauge invariant combination with the flux density
$F_{ij}$). The embeddings of the $STU$ model within the $N=8$ theory and
the consequent interpretation of its fields (scalar and vector) in
terms of dimensionally reduced 10 dimensional fields are achieved
using the powerful tool of solvable Lie algebras (SLA). This technique
will be briefly reviewed at the beginning of section 2. In the
subsections $2.1$ and $2.2$ the embeddings yielding NS--NS and R--R
charged black holes will be defined, together with the action of
$S\times T$ duality on the fields. The
mathematical details will be postponed to the appendix. In section 3
we shall consider, as an explicit example, a four parameter solution
of the $STU$ model, which can be easily characterized, at the
microscopic level, in terms of a bound state of D4 and D0--branes. 
The relation between macroscopic and microscopic parameters
will be particularly simple and immediate.

\section{The microscopic ``nature'' of the $STU$ model}

The $10$ dimensional interpretation of the fields characterizing a solution depends 
on the embedding of the $STU$ model inside the $N=8$ theory. An efficient technique for 
a detailed study of these embeddings is based on the so--called {\it Solvable Lie Algebra} 
(SLA) approach. In the following we shall summarize the main features of this 
formalism while we refer to \cite{mario} for a complete review on the subject. 

The solvable Lie algebra technique consists in defining a one to one correspondence between 
the scalar fields spanning a Riemannian homogeneous (symmetric) scalar manifold of the form 
${\cal M}= G/ H$ ($G$ being a non--compact semisimple Lie group and $H$ its maximal compact 
subgroup) and the generators of the  solvable subalgebra $Solv$ of the isometry 
algebra ${\cal G}$ defined by the well known Iwasawa decomposition:
\begin{eqnarray}
{\cal G}\,&=&\,{\cal H} \oplus Solv 
\label{iwa}
\end{eqnarray}
where ${\cal H}$ is the compact algebra generating $H$. A Lie algebra $ G_s$ is {\it solvable} 
if for some integer $n\geq 1$, its $n^{th}$ order {\it derived} algebra vanishes:
\begin{eqnarray}
{\cal D}^{(n)} G_{s}&=&0 \qquad {\mbox{where}} \nn \\
\cD G^{(1)}_s&\equiv&[ G_s, G_s]\quad;\quad \cD^{(k+1)} G_s \equiv [\cD^{(k)} G_s,
\cD^{(k)} G_s]\nonumber
\end{eqnarray}
Since the $70$--dimensional  scalar manifold ${\cal M}_{scal}$ of
$N=8$ supergravity  has the above coset structure with $G=E_{7(7)}$
and $H=SU(8)$, it can be globally  described as the group manifold
generated by a solvable Lie algebra $Solv_7$, whose  parameters are
the scalar fields $\phi_i$:  \beqa Solv_7=\{T_i\}\qquad \phi_i
\leftrightarrow T_i \quad i=1,\dots, 70\nn \eeqa Indeed the solvable
group generated by $Solv_7$ acts  transitively on ${\cal
M}_{scal}$. Considering the $N=8,\,d=4$ theory as the dimensional
reduction on a torus $T^6$ of type IIA or IIB supergravity theories in
$d=10$, the solvable characterization of the NS--NS and R--R scalars
in the four dimensional theory was worked out in \cite{solv,RR}
\footnote{In those papers the correspondence  generators--scalars
referred just to one of the two maximal theories. In the  present
paper we wish to give a precise geometrical characterization of the
bosonic  fields of the two theories and to find the relation between
them.} and is achieved by  decomposing the solvable algebra $Solv_7$
with respect to the solvable algebra  $Solv_T+Solv_S$, where $Solv_T$
generates the moduli space of the torus  ${\cal
M}_T=SO(6,6)/SO(6)\times SO(6)$ ($T=SO(6,6)$ being the classical
$T$--duality group), and $Solv_S$ generates the two dimensional
manifold $SL(2,\IR)/SO(2)$ spanned by the dilaton $\phi$ and the axion
$B_{\mu\nu}$  ($S=SL(2,\IR)$ being the $S$--duality group of the
classical theory). Since  in the formalism outlined above $Solv_T$ is
naturally parameterized by the moduli  scalars $G_{ij},\,B_{ij}$
($i,j$ denoting the directions inside the torus), and  $Solv_S$ by
$\phi$ and $B_{\mu\nu}$, the complement of $Solv_T+Solv_S$ inside
$Solv_7$  is a $32$--dimensional subspace of nilpotent generators
parameterized by the 32 R--R scalars
\footnote{Depending on whether the ${\bf 32}$ representation of
$SO(6,6)$ has positive or  negative chirality (i.e. ${\bf 32}^+$ and
${\bf 32}^-$, described within two different  constructions of the
${\cal E}_{7(7)}$ algebra), the corresponding theory will be type  IIA
or type IIB, as we shall see. }. The general structure of the solvable
algebra defined by the decomposition (\ref{iwa}) is the direct sum of
a  subspace of the Cartan subalgebra CSA and the nilpotent space
spanned by the  {\it shift} operators corresponding to roots whose
restriction to this Cartan subspace is positive:  \beq Solv\,=\,{\cal
C}_K \oplus \sum_{\alpha \in \Delta^+}\{E_{\alpha}\} \eeq ${\cal C}_K$
is the non--compact part of the CSA and $\Delta^+$ is the space of
those roots which are positive (non vanishing) with respect to ${\cal
C}_K$.

In the case of the $N=8 $ theory in $d=4$, $Solv$ is generated by the
generators of the  whole Cartan subalgebra of ${\cal E}_{7(7)}$ (which
denotes the algebra generating the  group $E_{7(7)}$, whose Cartan
generators are non--compact) and all the shift operators corresponding
to the positive roots  of the same algebra. A suitable basis of Cartan 
generators is parametrized by the {\it radii}
of the internal torus  $\rho_i$ 
(i.e. of the cycles along orthogonal compact directions $x^i$) plus the 
dilaton $\phi$;
 the shift operators corresponding to the positive roots of
$Solv_T$  are parametrized by the remaining $T^6$ moduli; finally, the
shift operators corresponding to the positive {\it spinorial} roots of
$Solv_T$ are naturally  parameterized by the R--R scalars. The precise
correspondence between the positive roots of $E_{7(7)}$ and type IIA
and type IIB fields is summarized in the appendix.  Although this
correspondence is fixed by the geometry, in what follows we shall
define algebraically two different classes of {\it embeddings} of the
$STU$ model  within the $N=8$ theory which describe NS--NS or R--R
charged solutions respectively. The  embeddings within each class are
related by an $S\times T$ transformation which preserves, as a general
property, the NS--NS and R--R nature of the fields. This
transformation is implemented on the scalar and vector fields  through
the action of the {\it automorphisms} ($Aut(S\times T)$) of the
$S\times T$ duality group $SL(2,\IR)\times SO(6,6)$  on the
corresponding generators and  weights, respectively. As far as the
action on the generators is concerned, the {\it inner automorphisms}
will amount to  a ``rotation'' of the Cartan generators (separately
acting on the radii and the dilaton) and a redefinition of the NS--NS
scalars $G_{ij}$ ($i\neq j$) and $B_{ij}$ and of the R--R fields
within the ${\bf 32}^+$ or  ${\bf 32}^-$ (type IIA or type IIB
respectively). The action of the {\it outer automorphisms} of $D_6$
(the Dynkin diagram of $SO(6,6)$) differs from the one described above
in the fact that it exchanges  the ${\bf 32}^+$ with the  ${\bf 32}^-$
weights of $SO(6,6)$ and thus the type IIA theory with type IIB
theory. This is consistent with the characterization of $T$--duality
along a compact direction which we shall give in the sequel.

Let us now recall the main concepts on how to define the embedding of
the $STU$ model  from the reduction of the central charge matrix
$Z_{AB}$ of the $N=8$ theory to its  skew--diagonal form ({\it normal form}), 
$Z^{{\cal N}}$:
\begin{equation}
Z^{\cal N}\, =\,\left( \begin{array}{clcl} Z_1\,\epsilon & 0 & 0 & 0
\\ 0 & Z_2\,\epsilon & 0 & 0 \\ 0 & 0 & Z_3\,\epsilon & 0 \\ 0 & 0 & 0
& Z_4\,\epsilon
\end{array}
\right) \qquad,\qquad \epsilon\,=\,\left( \begin{array}{cl} 0 & 1 \\
-1 & 0
\end{array}
\right)
\label{nform}
\end{equation}

The $U$--duality invariant properties of an $N=8$ BPS black hole solution
are represented by the $SU(8)$ invariants which may be built out of $Z_{AB}$.
There are $5$ invariants {\it intrinsically} associated with the
central charge matrix: the four norms of its skew--eigenvalues $Z_\alpha$ and the overall 
phase of the latter \cite{malser}. 
They represent the five duality--invariant quantities characterizing a $1/8$ 
BPS black hole (i.e. its essential macroscopic 
degrees of freedom). The skew--eigenvalues of the central charge matrix do not depend 
on $16$ of the $70$ scalar fields (which are associated with the {\it centralizer} of 
$Z^{{\cal N}}$ \cite{mp1}, to be defined in the sequel) while they depend on the $56$ 
quantized charges $\vec{Q}=(p,q)$ and the remaining $54$ scalars through the entries $Z_{AB}$.
Since we are interested in the generating solution, we may start looking for a 
minimal consistent truncation of the $N=8$ theory on which the four complex 
(eight real parameters) skew--eigenvalues of $Z_{AB}$ are independent parameters. 
This may be achieved by means of an $SU(8)$ gauge fixing which amounts to setting to 
zero all the central charge entries except the skew--diagonal 
ones. The above gauge fixing corresponds to a $48$ parameter $U$--duality transformation 
on the $56$ quantized charges and the $54$ scalars in the expression of the central charge 
skew--eigenvalues, which makes $Z^{{\cal N}}$ depend only on eight quantized charges 
$\vec{Q}^{\cal N}=(p^{\cal N},q^{\cal N})$ (the {\it normal form} of the quantized
charges) and on $6$ scalar fields defining in turn the vector and the
scalar content of an $STU$ model. The generating solution will be a solution
within this $STU$ model, depending only on {\it five} of the eight quantized charges
(resulting from fixing a residual $SO(2)^3$ gauge  transformations acting on  $Z^{{\cal N}}$).
After the $SU(8)$ gauge fixing the entries of  $Z^{{\cal N}}$ will coincide with the 
skew--diagonal entries of  $Z_{AB}$ and therefore the result of this procedure will 
depend on the initial basis in which the central charge matrix is written.

The embedding of the $STU$ model, resulting from the above gauge fixing, within the $N=8$ 
theory can be characterized geometrically as follows \cite{mp1},\cite{mp2}. We define the 
{\it centralizer} of $\vec{Q}^{{\cal N}}$ as its {\it little group} $G_C=SO(4,4)$ contained 
inside $E_{7(7)}$ (i.e.  $G_C\cdot \vec{Q}^{{\cal N}}=\vec{Q}^{{\cal N}}$). The maximal 
compact subgroup $H_C=SO(4)^2$ of $G_C$ is the centralizer of $Z^{{\cal N}}$ and the 
homogeneous manifold $G_C/H_C$ is spanned by the aforementioned $16$ scalar fields which 
the skew--eigenvalues of the central charge do not depend on. We define the {\it normalizer} 
of  $\vec{Q}^{{\cal N}}$ as the maximal subgroup $G_{\cal N}=\left[SL(2,\IR)\right]^3$ of 
$E_{7(7)}$ of which $\vec{Q}^{{\cal N}}$ is an irreducible representation (namely the 
${\bf (2,2,2)}$). As a general property the condition $\left[G_C,G_{\cal N}\right]=0$ 
must hold. The maximal compact subgroup $H_{\cal N}=\left[SO(2)\right]^3$ of  $G_{\cal N}$
is the normalizer of $Z^{{\cal N}}$ and the six dimensional homogeneous coset 
$G_{\cal N}/H_{\cal N}$ defines  the scalar manifold of the $STU$ model we are interested in:
\beqa
{\cal M}_{STU}\,=\,\left[\frac{SL(2,\IR)}{SO(2)}\right]^3
\label{mscal}
\eeqa 
The scalar content of this model, in terms of the $N=8$ scalars, is defined 
by embedding  $Solv({\cal M}_{STU})$ into $Solv_7$, $\vec{Q}^{{\cal N}}$ defines 
the quantized charges of the model, while as usual the real and imaginary parts of
the skew eigenvalues $Z_\alpha$ of the central charge define the physical
dressed electric and magnetic charges of the interacting $N=2$ model.

As previously stressed, the  above defined $SU(8)$ gauge fixing procedure, 
when applied to $Z_{AB}$ in {\it different} bases, yields $STU$
models embedded {\it differently} inside the original theory (the embedding of the algebras 
${\cal G}_{\cal N}$ and ${\cal G}_C$, generating $G_{\cal N}$ and $G_{C}$, inside 
${\cal E}_{7(7)}$ would in general depend on the original basis of $Z_{AB}$). 
In the following subsections we shall define two relevant classes of 
embeddings of the $STU$ model in the $N=8$ theory, in which the vector fields have a NS--NS 
or a R--R ten dimensional origin respectively, once the latter theory is interpreted as 
the low energy supergravity of a type II string on $T^6$. 

\subsection{The NS--NS $STU$ model}

Let us consider the central charge matrix in a basis
$Z_{\hat{A}\hat{B}}$ in which the index $\hat{A}$ of the ${\bf 8}$ of
$SU(8)$  splits in the following way:
$\hat{A}=(a=1,\dots,4;a^\prime=1^\prime\dots,4^\prime)$,  where $a$
and $a^\prime$ index the ${\bf (4,1)}$ and  ${\bf (1,4^\prime)}$ in
the decomposition of the ${\bf 8}$ with respect to $SU(4)\times
SU(4)^\prime=SU(8)\cap SO(6,6)$  (this is the basis considered by
Cvetic and Hull in defining their NS--NS $5$--parameter  solution,
\cite{hull}). The group $SU(4)\times SU(4)^\prime$ is the maximal
compact subgroup of the classical $T$--duality group and decomposing
with respect to it the ${\bf 28}$ of $SU(8)$ will define which of the
entries of $Z_{\hat{A}\hat{B}}$ correspond to R--R and which to NS--NS
vectors (the former will transform in the spinorial of $SU(4)^2\equiv
SO(6)^2$):
\begin{eqnarray}
{\bf 28}\rightarrow \bf{(6,1^\prime)+(1,6^\prime)+(4,4^\prime)}
\label{Zrrns}
\end{eqnarray}     
the $\bf{(6,1^\prime)+(1,6^\prime)}$ part consists of the two diagonal
blocks $Z_{ab}$ and $Z_{a^\prime b^\prime}$ and define the $12$ NS--NS
(complex) charges,  while the spinorial ${\bf (4,4^\prime)}$
correspond to the off--diagonal block $Z_{aa^\prime}$ and  define the
$16$ (complex) R--R charges. The skew--diagonal elements which will
define $Z^{{\cal N}}_{NS}$ correspond then to NS--NS charges
($Z_{12},Z_{34},Z_{1^\prime 2^\prime}, Z_{3^\prime 4^\prime}$) and
therefore the corresponding $STU$ model will contain $4$ NS--NS vector
fields. Let us work out the embedding of $G_{\cal N}$ and $G_C$ within
$E_{7(7)}$. Let the simple roots of  ${\cal E}_{7(7)}^\pm$\footnote{
Let us recall that the Dynkin diagram of  ${\cal E}_{7(7)} $ is
constructed by adding to the $D_6$ Dynkin diagram  consisting of the
simple roots $(\alpha_i)_{i=1,\dots,6}$ the highest weight
$\alpha_7^\pm$ of one of the spinorial representations ${\bf 32}^\pm$
of $SO(12)$.  We call ${\cal E}_{7(7)}^+$ the algebra obtained by
attaching $\alpha_7^+$ to  $\alpha_5$; ${\cal E}_{7(7)}^-$ the algebra
obtained by attaching $\alpha_7^-$  to $\alpha_6$. As previously
anticipated the scalars of type IIA or type IIB  theories parameterize
respectively $Solv({\cal E}_{7(7)}^+)$ or $Solv({\cal E}_{7(7)}^-)$.}
be $\alpha_n$ whose expression with respect to an orthonormal  basis
$\epsilon_n$ is the following:
\begin{eqnarray}
\alpha_1\,&=&\,\epsilon_1-\epsilon_2\,\,;\,\,\alpha_2\,=\,\epsilon_2-\epsilon_3\,\,;
\,\,\alpha_3\,=\,\epsilon_3-\epsilon_4\nonumber\\
\alpha_4\,&=&\,\epsilon_4-\epsilon_5\,\,;\,\,\alpha_5\,=\,\epsilon_5-\epsilon_6\,\,;
\,\,\alpha_6\,=\,\epsilon_5+\epsilon_6\nonumber\\ \alpha_7^\pm\,&=&\,-
\frac{1}{2}\left(\epsilon_1+\epsilon_2+\epsilon_3+\epsilon_4+
\epsilon_5\mp\epsilon_6\right)+\frac{\sqrt{2}}{2}\epsilon_7
\label{dynke7pm}
\end{eqnarray}
As previously stated, in the SLA formalism the Cartan subalgebra is 
parametrized by the six radii of the torus and the dilaton, in 
particular the orthonormal elements $H_{\epsilon_i}$ are multiplied by
the scalars ${\rm ln}(\rho_{i+3})$ ($i=1,\dots,6$), see (\ref{strange}).

The group $H_C=SO(4)^2\subset SO(6)^2\subset
SO(6,6)$ consists of four $SU(2)$ factors acting separately on the
blocks $(1,2)$, $(3,4)$, $(1^\prime 2^\prime)$, $(3^\prime 4^\prime)$
of the central  charge matrix. The centralizer at the level of
quantized charges $G_C$ on the other hand is the group $SO(4,4)$
regularly embedded in $SO(6,6)$. If the latter is described by the
simple roots $\alpha_1,\dots,\alpha_6$, a simple  choice, modulo
$S\times T$--duality transformations, for the Dynkin diagram of ${\cal
G}_C$ would be $\alpha_3,\alpha_4,\alpha_5,\alpha_6$. The solvable
subalgebra of ${\cal G}_C$ consists of NS--NS generators only. The
algebra ${\cal G}_{\cal N}$, being characterized  as the largest subalgebra
of $Solv_7$ which commutes with ${\cal G}_C$, is immediately  defined,
modulo isomorphisms, to be the $\left[SL(2,\IR)\right]^3$ algebra
corresponding to the roots $\beta_1=\sqrt{2}\epsilon_7$,
$\beta_2=\epsilon_1-\epsilon_2$  and
$\beta_3=\epsilon_1+\epsilon_2$. The scalar manifold of the
corresponding  $STU$ model has the form:
\begin{eqnarray}
{\cal M}_{STU}\,&=&\,\frac{G_{\cal N}}{H_{\cal
N}}=\frac{SU(1,1)}{U(1)}(\beta_1)\times \frac{SO(2,2)}{SO(2)\times
SO(2)}(\beta_2,\beta_3)
\end{eqnarray}
The reason why the above expression has been written in a factorized
form is to stress  the different meaning of the two factors from the
string point of view: the group  $SU(1,1)(\beta_1)$ represents the
classical $S$--duality group of the theory and the corresponding
factor of the manifold is parameterized by the dilaton $\phi$ and the
axion $B_{\mu\nu}$.  In the same way it can be shown that the  second
factor is parameterized by the scalars  $G_{44}\,,\,G_{55}\,,\,G_{45}$
and $B_{45}$ and its isometry group acts as a classical  $T$--duality,
i.e. its restriction to the integers is the perturbative $T$--duality
of string theory. This non--symmetric version of the $STU$ model is
the same as the one obtained as a consistent truncation of the
toroidally compactified heterotic  theory and therefore describes the
generating solution also for this theory (the string interpretation of
the $4$ scalars spanning the second factor  in ${\cal M}_{STU}$ is in
general non generalizable to the heterotic theory). Therefore its
corresponding microscopic  structure should be  given in terms of NS
states (fundamental string and NS5--brane states).
  
\subsection{The R--R $STU$ model}

Let us start with the central charge matrix $Z_{AB}$ obtained from
$Z_{\hat{A}\hat{B}}$ through an orthogonal conjugation, such that the
new index $A$ of the ${\bf 8}$ of $SU(8)$ assumes the values
$A=1,1^\prime,2,2^\prime,\dots,4,4^\prime$, the unprimed and primed
indices  spanning the ${\bf 4}$ of the two $SU(4)$ subgroups
previously defined.  Let us now consider the decomposition of $SU(8)$
with respect to its subgroup  $U(1)\times SU(2)\times SU(6)$ (which is
the decomposition suggested by the Killing  spinor analysis of the
$1/8$ BPS black holes) such that the ${\bf 8}$ decomposes into  a
${\bf (1,2,1)}$ labeled by $i=4,4^\prime$ and a ${\bf (1,1,6)}$
labeled by  $\tilde{i}=1,1^\prime,\dots,3,3^\prime$. The ${\bf 28}$
decomposes with respect to  $U(1)\times SU(2)\times SU(6) $ in the
following way:
\begin{eqnarray}
{\bf 28}\rightarrow {\bf (1,1,1)+(1,1,15)+(1,2,6)}
\label{dec126}
\end{eqnarray}
where the singlet represents the diagonal block $Z_{ij}$, the ${\bf
(1,1,15)}$ the  diagonal block $Z_{\tilde{i},\tilde{j}}$ and the ${\bf
(1,2,6)}$ is spanned by the off diagonal entries
$Z_{i,\tilde{j}}$. The skew--diagonal entries which survive the previously 
defined gauge
fixing procedure and which thus enter the new normal form
of the central charge $Z_{RR}^{{\cal N}}$ are now
$Z_1=Z_{1,1^\prime}$,$Z_2=Z_{2,2^\prime}$,$Z_3=Z_{3,3^\prime}$ and
$Z_4=Z_{4,4^\prime}$, which are R--R charges.\par It is
interesting to notice that these four (complex) charges are part of
the set of $10$ R--R (complex) charges entering the diagonal blocks
${\bf (1,1,1)+(1,1,15)}$.  These charges can be immediately worked out either
by directly counting the entries with mixed primed and unprimed indices
($Z_{ab^\prime}$) contained in these two blocks or, in a group
theoretical fashion, by decomposing the ${\bf (1,1,1)+(1,1,15)}$ in
(\ref{dec126}) and the ${\bf (4,4^\prime)}$  in (\ref{Zrrns})  with
respect to a common subgroup $U(1)\times SU(3)\times
SU(3)^\prime\equiv \left[U(1)\times SU(6)\right] \cap \left[SU(4)\times
SU(4)^\prime\right]$. Both the decompositions  contain a common
representation ${\bf (1,1,1^\prime)+(1,3,3^\prime)}$  describing $10$
R--R central charges. The ${\bf 3}$ and ${\bf 3^\prime}$ are  spanned
by the values $1,2,3$ and $1^\prime,2^\prime,3^\prime$ of the indices
$a$ and $a^\prime$ of the ${\bf 4}$ and ${\bf 4^\prime}$ respectively.
These charges correspond to the $1+9$ vectors of an $N=2$ truncation
of the  $N=8$ theory with scalar manifold $SU(3,3)/U(3)\times SU(3)$.
A truncation of this theory yields in turn the  $STU$ model 
corresponding to the normalizer of $Z_{RR}^{{\cal N}}$. 
The $4$ complex charges in
$Z_{RR}^{{\cal N}}$ will indeed depend on the $8$ R--R quantized magnetic
and electric  charges $\vec{Q}^{{\cal N}}_{RR}$ and the $6$ scalar
fields of the new $STU$ model.  Therefore, differently to the previous
defined class, in this case all gauge fields  (and hence the
corresponding charges) come from R--R 10 dimensional forms.  The
centralizer $SO(4,4)$ of $\vec{Q}^{{\cal N}}_{RR}$ is now no more
contained inside  $SO(6,6)$ and therefore its solvable algebra
contains R--R generators as well.  As a common feature of the
truncations belonging to this class, the scalars  entering each
quaternionic multiplet split into $2$ NS--NS and $2$ R--R. Indeed the
centralizer $SO(4,4)$ is now the isometry group of the manifold
$SO(4,4)/SO(4)\times SO(4)$ describing $16$ hyperscalars and therefore
its solvable algebra has $8$ R--R and $8$ NS--NS generators
\footnote{Notice that any solution within a Calabi--Yau
compactification of type II string lies in this class. Indeed all the
vector fields surviving the compactification come from R--R forms,
both for type IIA and type IIB theories.}.

In order to specify a particular truncation within the class one
should define the simple roots  of $ SO(4,4)$ and of the isometry
group $\left[SL(2,\IR)\right]^3$ of the $STU$ model, which in turn
determines $Solv({\cal M}_{STU})$ and thus the scalar content of the
model.  An interesting possibility is the one where the system of
simple roots for $SO(4,4)$ is chosen to be:
\begin{eqnarray}
\gamma_1\,&=&\,\epsilon_1+\epsilon_2\,\,;\,\,\gamma_3\,=\,\epsilon_3+\epsilon_4\,\,;
\,\,\gamma_4\,=\,\epsilon_5+\epsilon_6\nonumber\\
\gamma_2\,&=&\,\alpha^-_7
\label{gammas} 
\end{eqnarray} 
Here the root
$\beta_1=\sqrt{2}\epsilon_7=\sum_{i=1}^{4}\gamma_i+2 \gamma_2$ belongs
to the $SO(4,4)$ root space. In the solvable language, since the
Cartan generator and the shift operator corresponding to this root are
parameterized by $\phi$ and $B_{\mu\nu}$ respectively, these two
scalars are now part of a hypermultiplet, known as the {\it universal
sector}. The isometry group of the $STU$ model which commutes with the
above defined $SO(4,4)$ centralizer is generated by a
$\left[SL(2,\IR)\right]^3$ algebra which is regularly embedded in the
isometry group $GL(6,\IR)$ of the classical moduli space of $T^6$ and
defined by the following roots:
\begin{eqnarray}
\beta_1\,&=&\,\epsilon_1-\epsilon_2\,\,;\,\,\beta_2\,=\,\epsilon_3-\epsilon_4\,\,;
\,\,\beta_3\,=\,\epsilon_5-\epsilon_6
\label{beta} 
\end{eqnarray} 
The scalar manifold of this $STU$ model is now symmetric with respect to 
 $S,T,U$ since it is contained in the moduli space of $T^6$ (its scalars are
all NS--NS but there is no $\phi$ and $B_{\mu\nu}$):  \begin{eqnarray}
\label{stusym} {\cal M}_{STU}\,&=&\,\frac{G_{\cal N}}{H_{\cal
N}}=\frac{SU(1,1)}{U(1)}(\beta_1)\times
\frac{SU(1,1)}{U(1)}(\beta_2)\times \frac{SU(1,1)}{U(1)}(\beta_3)
\label{mrr}
\end{eqnarray} 
>From table \ref{solfil} we can read out the scalar content of this
model: $G_{45}\,,\,G_{67}\,,\,G_{8\,9}$ and $3$ radii (the latter
being Cartan generators). The interesting feature of the above
embedding is that all the excited scalar fields come from the metric
tensor $G$ rather than from the Kalb--Ramond field $B$ (on the
contrary, and this is a common feature of all embeddings falling in
this class, all charges are R--R).  Suppose we are working in the
framework of type IIB theory,  then this particular embedding could
possibly be described in the weak string coupling  regime by a system
of D3--branes at angles.

Let us denote the compact directions of the torus are
$x^{4,5,6,7,8,9}$ and the non--compact space--time coordinates are
$x^{0,1,2,3}$.  It is interesting to consider an other embedding which
is obtained from the one above by acting on it by means of a
$T$--duality along the directions $x^5,x^7,x^9$ of the internal
torus. In the SLA language this operation is characterized by three
transformations $\psi_{\tau_1},\,\psi_{\tau_2},\,\psi_{\tau_3}$ in 
$Aut(S\times T)$ (see the appendix), where $\tau_i$ are rotations on the root space obtained by multiplying the outer automorphism of $D_6$
($\alpha_5\leftrightarrow \alpha_6$) with suitable Weyl
transformations on $D_6$. The automorphisms $\psi_{\tau_i}$ act on the Cartan
subalgebra by sending $H_{\epsilon_{2,4,6}}\rightarrow
-H_{\epsilon_{2,4,6}}$ and therefore the corresponding fields
transform in the following way: $\rho_k\rightarrow
\rho_k^{-1},\,\,(k=5,7,9)$.  The action of the automorphism
$\psi_{\tau}=\psi_{\tau_1\tau_2\tau_3}$ may be extended from the Cartan subalgebra to the shift operators $E_\alpha$, 
and therefore on the corresponding axions, using the recipe (\ref{simplerecipe}): $\psi_{\tau}(E_\alpha)\propto E_{\tau(\alpha)}$. Since
\begin{eqnarray}
&&\tau(\epsilon_1-\epsilon_2)=\epsilon_1+\epsilon_2\nonumber\\
&&\tau(\epsilon_3-\epsilon_4)=\epsilon_3+\epsilon_4\nonumber\\
&&\tau(\epsilon_5-\epsilon_6)=\epsilon_5+\epsilon_6\nonumber
\end{eqnarray}
 the corresponding axions will transform as follows (see appendix):
\begin{eqnarray}
&&\psi_\tau :\cases{G_{45}\rightarrow B_{45}\cr G_{67}\rightarrow B_{67}\cr
 G_{89}\rightarrow B_{89}}
\end{eqnarray} 

A $T$-duality along an odd number of internal directions, which in our
 characterization amounts to inverting the sign to an odd number of
 $\epsilon_i$, maps ${\bf 32}^-$ into ${\bf 32}^+$ (see the appendix)
 that is type IIB theory into type IIA and this is consistent with
 known properties of $T$--duality. This new embedding
 $SL(2,\IR)^3\subset SU(3,3)\subset {\cal E}_{7(7)}^+$, obtained by
 acting with the $T$--duality $\tau$ in on the embedding of
 $SL(2,\IR)^3$ in ${\cal E}_{7(7)}^-$ defined by eqs. (\ref{beta}) and
 (\ref{mrr}), is a type IIA embedding in which the axions come only
 from internal components of the antisymmetric tensor $B_{ij}$. It is
 natural to interpret black hole solutions within this embedding, for
 instance, in terms of systems of D0 and D4 branes with  magnetic flux
 on the world volume of the latter, as anticipated in the
 introduction.  The mathematical details associated with the two R--R
 embeddings described above (i.e. the one in type IIB and the
 $T$--dual in type IIA theory) will be dealt with in the appendix
 . These two embeddings provide a convenient framework in which to
 look for the generating solution and to give it a suitable D--brane
 interpretation in which a microscopic entropy counting is affordable.
 In the next section we shall fix on the type IIA R--R embedding and
 consider, just as an example, a known four parameter solution, which
 is a pure dilatonic one, and  work out both its macroscopic and
 microscopic description, consistently with the 10 dimensional
 interpretation of its scalar and vector fields given in the appendix.

\section{Example: a pure R--R solution and its microscopic description}

Let us now consider a specific example, namely a four parameter
solution within the  $STU$ model. From  the macroscopic point of view
this solution is analogous to the one described in \cite{ferka}.
Other macroscopic solutions of the $STU$ model have been obtained, for
instance, in  \cite{kal1,sabra1,bft,bft2}.

Let us very briefly remind the structure of the $STU$ model while a
complete treatment has been  carried out in \cite{bft,bft2}. The $STU$
model is characterized by  a $N=2$ supergravity theory coupled to 3
vector multiplets whose  scalars spans the manifold  ${\cal M}_{STU}$,
eq.(\ref{mscal}). The total number of scalar fields in the game is $6$
($z_i=a_i+ib_i\,,\,i=1,2,3$) while the number of charges
$(p^\Lambda,q_\Lambda)$ is  $8$ ($4$ electric and $4$ magnetic). In
the framework of the $STU$ model, the local realization  on moduli
space ${\cal M}_{STU}$ of the $N=2$ supersymmetry algebra central
charge $Z$ and of  the $3$ {\it matter} central charges $Z^i$
associated with the $3$ matter vector fields are  related to the $N=8$
central charge eigenvalues in the following way: \beqa
\label{stun8}
Z\,=\,i\,Z_4 \quad,\quad
Z^i\,=\,h^{ij^\star}\nabla_{j^\star}\bar{Z}\,=\,\IP^i_{\hat{i}}
 Z^{\hat{i}} \eeqa
where $\IP^i_{\hat{i}}= 2 b_i(r)$ is the vielbein transforming rigid
indices $\hat{i}$ (the one  characterizing the eigenvalues of the $N=8$
central charge in its normal form, eq.(\ref{nform}))  to curved
indices $i$ (see \cite{bft}, section 3, for details).

The killing spinor equations characterizing the BPS black hole
solution translate into first  order differential equations for the
relevant bosonic fields once suitable ans\"atze are adopted.  As a
standard procedure, the vanishing of the  gravitino transformation
rule along killing spinor directions implies a condition for the
metric  while the vanishing of the dilatino transformation rules
translates into equations for the scalar  fields. The ans\"atze for
the metric $G_{\mu \nu}$ and the complex scalars $z^i$ are the
following:
\begin{eqnarray}
ds^2\,&=&\,e^{2{\cal U}\left(r\right)}dt^2-e^{-2{\cal
U}\left(r\right)}d\vec{x}^2 ~~~~~
\left(r^2=\vec{x}^2\right)\nonumber\\ z^i\,&\equiv &\, z^i(r)
\label{fishlicense}
\end{eqnarray}
After some algebra one can see that the structure of the first order
BPS equations turns out to  be the following:
\begin{eqnarray}
\frac{dz^i}{dr}\, &=&\, \mp 2\left(\frac{e^{{\cal U}(r)}}{r^2}\right)
h^{ij^\star}\partial_{j^\star} \vert Z(z,\bar{z},{p},{q})\vert \nn\\
\frac{d{\cal U}}{dr}\, &=&\, \mp \left(\frac{e^{{\cal
U}(r)}}{r^2}\right)  \vert Z(z,\bar{z},{p},{q})\vert
\label{eqs122}
\end{eqnarray}
which is a system of first order differential equations. Working out
the geometric  structure of the $STU$ model, the above
system of equations can be made explicit in terms of  the scalar
fields and the quantized charges $(p^\Lambda,q_\Lambda)$
characterizing the model.  This has been explicitly achieved in
\cite{bft,bft2} (where the same conventions and notations  have been
used), see in particular the appendices for explicit formul\ae.

In this section we shall focus a particular ($4$--parameter) regular
solution for which the  microscopic description turns out to be
particularly nice.  On this solution the central charge eigenvalues
$\{Z_\alpha\}\equiv\{Z_{\hat{i}}, Z_4\}$ are pure imaginary.  This condition fixes not only the
$\left[SO(2)\right]^3$ symmetry of the model but also the  overall
phase $\phi$ of the four central charge eigenvalues
($\phi=0\,\mbox{mod}\, 2\pi$),  yielding just four independent
invariants $\vert Z_\alpha\vert$.

Since the central charge $Z_4$ is set to be imaginary (i.e. $Z$ real),
the system of eqs.  (\ref{eqs122}) may be rewritten in the simpler
form:
\begin{eqnarray}
\frac{dz^i}{dr}\, &=&\, \mp \left(\frac{e^{{\cal U}(r)}}{r^2}\right)
h^{ij^\star}\nabla_{j^\star} \bar{Z}(z,\bar{z},{p},{q})\,=\, \mp
\left(\frac{e^{{\cal U}(r)}}{r^2}\right)Z^i(z,\bar{z},{p},{q}) \nn\\
\frac{d{\cal U}}{dr}\, &=&\, \mp \left(\frac{e^{{\cal
U}(r)}}{r^2}\right) Z(z,\bar{z},{p},{q})
\end{eqnarray} 
It is possible moreover to show that the reality of $Z$ is consistent
with the regularity of the  solution provided we set $p^0=0$.  The
conditions  $Z_{\hat{i}}=-\bar{Z}_{\hat{i}}$ (and therefore
$Z^i=-\bar{Z}^i$, see eq.(\ref{stun8})) imply that the three axions
are double--fixed: $a_{1,2,3}(r)\equiv a^f_{1,2,3}$.  They require
also that three electric quantized charges vanish, namely:
$q_1=q_2=q_3=0$.
 
Hence the quantized charges left are ($q_0,p^1,p^2,p^3$) and the
system of first order  differential equations our solution has to
fulfill reduces  considerably. Indeed the equations for the dilatons
and for ${\cal U}$ decouple from the axions  and may be solved
independently:
\begin{eqnarray}
\label{bfirst}
\frac{db_1}{dr}\,&=&\,\pm \left(\frac{e^{\cal U}}{r^2}
\right)\sqrt{-\frac{b_1}{2 b_2b_3}} \,(p^1b_2b_3-p^2 b_1b_3-p^3 b_1
b_2+q_0)\nonumber\\ \frac{db_2}{dr}\,&=&\,\pm \left(\frac{e^{\cal
U}}{r^2} \right)\sqrt{-\frac{b_2}{2 b_1b_3}} \,(-p^1b_2b_3+p^2
b_1b_3-p^3 b_1 b_2+q_0)\nonumber\\ \frac{db_3}{dr}\,&=&\,\pm
\left(\frac{e^{\cal U}}{r^2} \right)\sqrt{-\frac{b_3}{2 b_1b_2}}
\,(-p^1b_2b_3-p^2 b_1b_3+p^3 b_1 b_2+q_0)\nonumber\\ \frac{d{\cal
U}}{dr}\,&=&\,\pm \left(\frac{e^{{\cal U}}}{r^2} \right)
\frac{1}{2\sqrt{2}\sqrt{-b_1b_2b_3}}\,(p^1 b_2 b_3+p^2
b_1b_3+p^3b_1b_2+q_0)
\end{eqnarray}
The 3 equations for the axions and the one on the reality of the
central charge give the 4  $r$--independent relations our solution
should fulfill: \beqa \frac{da_1}{dr}&=& 0\,=\,
-\left({a_3}\,{b_2}+{a_2}\,{b_3}\right)\,{p^1} +  \left({a_3}\,{b_1} -
{a_1}\,{b_3}\right) \,{p^2} + \left( {a_2}\,{b_1} - {a_1}\,{b_2}
\right)\,{p^3}\nn\\ \frac{da_2}{dr}&=& 0\,=\,-\left( {a_3}\,{b_1} +
{a_1}\,{b_3} \right) \,{p^2} +  \left( {a_3}\,{b_2} - {a_2}\,{b_3}
\right)\,{p^1} + \left( {a_1}\,{b_2} - {a_2}\,{b_1}  \right)\,
{p^3}\nn\\ \frac{da_3}{dr}&=& 0\,=\,-\left( {a_1}\,{b_2} +
{a_2}\,{b_1} \right) \,{p^3} +  \left( {a_1}\,{b_3} - {a_3}\,{b_1}
\right) \,{p^2} + \left( {a_2}\,{b_3} - {a_3}\,{b_2}  \right) \,
{p^1}\nn\\ Im\,{\mbox Z}&=&0\,=\,\left( {a_3}\,{b_2} + {a_2}\,{b_3}
\right) \, {p^1} + \left( {a_3}\,{b_1} + {a_1}\,{b_3} \right) \, {p^2}
+ \left( {a_2}\,{b_1} + {a_1}\,{b_2} \right) \, {p^3} \eeqa The fixed
values for the scalar fields (namely the values the scalars get at the
horizon, \cite{fer}) are: \beqa &&b^{fix}_1\,=\,
-\sqrt{\frac{q_0p^1}{p^2p^3}}\;\;,\;\; b^{fix}_2\,=\,
-\sqrt{\frac{q_0p^2}{p^1p^3}}\;\;,\;\; b^{fix}_3\,=\,
-\sqrt{\frac{q_0p^3}{p^1p^2}} \nn \\ &&a_1^{fix}\,=\,0\quad,\quad
a_2^{fix}\,=\,0\quad,\quad a_3^{fix}\,=\,0 \eeqa Introducing four
harmonic functions as follows:
\begin{eqnarray}
H_{I}(r)\,&=&\,A_{I}+k_{I}/r
\,\,(I=0,1,2,3)\nonumber\\ k_{0}&=&\sqrt{2}\,q_0\quad,\quad
k_{i}= \sqrt{2}\,p^i
\end{eqnarray}
it is easy to see that the following ans\"atzefor the $b_i$ and the
scalar function  ${\cal U}$:
\begin{eqnarray}
b_1\,=\, -\sqrt{\frac{H_0H_1}{H_2H_3}}\;\;,\,\;\; b_2\,=\,
-\sqrt{\frac{H_0H_2}{H_1H_3}}\;\;,\,\;\; b_3\,=\,
-\sqrt{\frac{H_0H_3}{H_1H_2}}\;\;,\,\;\; {\cal
U}\,=\,-\frac{1}{4}\ln\,(H_0H_1H_2H_3)\nonumber \\
\label{ans}
\end{eqnarray}
satisfies both the first and second order differential
equations. Choosing the metric to be  asymptotically flat and standard
values for the dilatons at infinity, the four constants $A_I$ are
set to be all equal to $1$. The solution, consisting of the three
$b_i$, the double--fixed  $a_i$ and ${\cal U}$ is expressed in terms
of $4$ independent charges (and four harmonic functions):
$q_0,p^1,p^2,p^3$. According to the ans\"atze (\ref{fishlicense}) the
metric has the following form: \beqa ds^2\,&=&\,\left( H_0H_1H_2H_3
\right)^{-1/2}dt^2-\left( H_0H_1H_2H_3 \right)^{1/2} d\vec{x}^2  \eeqa
and the macroscopic entropy, according to Beckenstein--Hawking
formula, reads: \beq
\label{entrqp}
S_{macro}= 2\,\pi\,\sqrt{q_0p^1p^2p^3}  
\eeq 
As far as the vector 
fields are concerned, their form in terms of the harmonic functions
introduced  above is analogous to the one in the solution of
\cite{ferka} and we shall not give it here.  Let us now move to the
microscopic description of the above solution. This will be easily
obtained  starting from the explicit expression of the $N=8$ central
charge eigenvalues. Comparing eq.s  (\ref{eqs122}) and (\ref{bfirst})
one can see that the central charge eigenvalues at spatial  infinity,
when written in terms of the quantized charges, reads (in rigid
indices):
\begin{eqnarray}
Z_1\,&=&\,\frac{i}{2\sqrt{2}}(q_0 - p^2 - p^3 + p^1)\nonumber\\
Z_2\,&=&\,\frac{i}{2\sqrt{2}}(q_0 + p^2 - p^3 - p^1)\nonumber\\
Z_3\,&=&\,\frac{i}{2\sqrt{2}}(q_0 - p^2 + p^3 - p^1)\nonumber\\
Z_4\,&=&\,\frac{i}{2\sqrt{2}}(q_0 + p^2 + p^3 + p^1)
\label{chargeqp}
\end{eqnarray}
According to the previous discussion, if the $STU$ model is embedded
in the full $N=8$  theory as discussed in subsection 2.2, the
microscopic  description of the above solution can be given in terms
of the intersection of four bunches of parallel  D--branes. In
particular, if we consider the type IIA embedding defined in the same
subsection, the  central charge $Z_4$ (which represents the $N=2$
graviphoton dressed charge) and the matter charges  $Z_{\hat{i}}$
 are related to the gauge fields coming from the $10$
dimensional  R--R 3--form $A_{MNP}$ (and its Hodge dual) coupled to
D$2$ (and D$4$)--branes and from the R--R 1--form $A_M$ (and its Hodge
dual)  coupled to D$0$ (and D$6$)--branes. Our solution is hence
described,  at the microscopic level, as a $1/8$ supersymmetry
preserving intersection of $4$ bunches of  these D--branes. The fact
that each of the central charge eigenvalues is real or pure imaginary
(in our case they are all pure imaginary) implies that the solution is
pure electric,  that is it is not made of electromagnetic dual objects.

Let us be more precise and try to deduce the microscopic
interpretation of our solution from the algebraic framework
consistently built in \cite{mp1},\cite{bft} and in the present paper
(see the appendix). In appendix A of \cite{bft}  the $Sp(8,\IR)$
representation of the $SL(2,\IR)^3$ generators that we have been using
for our study of the $STU$ model is given. In particular, the Cartan
generators have the form\footnote{After a suitable change of basis by 
means of the matrix M defined in \cite{bft}.}:
\begin{eqnarray}
h_1&=&\frac{1}{2}{\rm diag}\left(
-1,1,-1,-1,1,-1,1,1\right)\nonumber\\ h_2&=&\frac{1}{2}{\rm
diag}\left( -1,-1,1,-1,1,1,-1,1\right)\nonumber\\
h_3&=&\frac{1}{2}{\rm diag}\left( -1,-1,-1,1,1,1,1,-1\right)
\label{hi}
\end{eqnarray} 
If we interpret these Cartan generators as corresponding (in the type
IIA framework) to the roots $\epsilon_1+\epsilon_2$,
$\epsilon_3+\epsilon_4$ and $\epsilon_5+\epsilon_6$, so that we can
interpret their diagonal values as the scalar product of these roots
with 8 weights ${W^{(\lambda)}}^-$, the only set of weights on which
all the three generators are non singular as in (\ref{hi})
\footnote{Which is a general feature of the $Sp(8)_D$ representation
of $SL(2,\IR)^3$.} is:
\begin{eqnarray}
\{{W^{(1)}}^-,{W^{(6)}}^-,{W^{(7)}}^-,{W^{(16)}}^-,{W^{(29)}}^-,{W^{(34)}}^-,{W^{(35)}}^-,
{W^{(44)}}^-\}
\end{eqnarray}
A possible correspondence is:
\begin{eqnarray}
h_1&=&\frac{-1}{2}H_{\epsilon_1+\epsilon_2}\nonumber\\
h_2&=&\frac{-1}{2}H_{\epsilon_3+\epsilon_4}\nonumber\\
h_3&=&\frac{-1}{2}H_{\epsilon_5+\epsilon_6}\nonumber\\
\{p^0,p^1,p^2,p^3\}&=& \{{W^{(29)}}^-,{W^{(7)}}^-,{W^{(16)}}^-,{W^{(6)}}^-\}\nonumber\\
\{q_0,q_1,q_2,q_3\}&=&\{{W^{(1)}}^-,{W^{(35)}}^-,{W^{(44)}}^-,{W^{(34)}}^-\}
\label{iden}
\end{eqnarray} 
Since our solution is charged only with respect to $\{q_0,p^1,p^2,p^3\}$ from table 3 we 
may read off the corresponding R--R fields: $A_\mu,A_{\mu 6789},A_{\mu 4589},A_{\mu 4567}$. 
The configuration of microscopic objects coupled with these forms consists in $3$ bunches 
of orthogonal D4--branes ($N_1\,,\,N_2\,,\,N_3$, respectively) wrapped on the internal 
torus $T^6$ with $N_0$ D0--branes on top of them where the D4--branes are positioned in the 
following way:

\begin{table} [ht] 
\begin{center}
\begin{tabular}{|c|c|c|c|c|c|c|}
\hline
  & $x^4$ & $x^5$ & $x^6$ & $x^7$ & $x^8$ & $x^9$ \\
\hline
$N_1$ & $\cdot$  & $\cdot$  & $\times$  & $\times$  & $\times$ & $\times$ \\
\hline
$N_2$ & $\times$  & $\times$  & $\cdot$  & $\cdot$  & $\times$ & $\times$ \\
\hline
$N_3$ & $\times$  & $\times$  & $\times$  & $\times$  & $\cdot$ & $\cdot$ \\
\hline
\end{tabular}
\end{center}
\caption{{\small The position of the $D4$ branes on the compactifying torus: for any given brane 
the directions labeled with $\times$ are Neumann while those labeled with $\cdot$ are 
Dirichlet.}}
\label{NDbrane}
\end{table}

The above configuration is $1/8$ supersymmetric and adding any number of $D0$ branes the number 
of preserved supersymmetries does not change, \cite{jab}. The previous analysis about the 
microscopic interpretation of the vector fields suggests a precise relation between 
$\{q_0,p^1,p^2,p^3\}$ and $\{N_0,N_1,N_2,N_3\}$ in the order. This relation can be easily 
derived by writing the expression of the $E_{7(7)}$ quartic invariant, $J_4$, as in 
\cite{malser}:
\beqa
J_4&=&(|Z_1|+|Z_2|+|Z_3|+|Z_4|)\,(|Z_1|-|Z_2|-|Z_3|+|Z_4|)\,(-|Z_1|+|Z_2|-|Z_3|+|Z_4|)\nn \\
&&(-|Z_1|-|Z_2|+|Z_3|+|Z_4|) + 8 |Z_1||Z_2||Z_3||Z_4|\,(\cos \phi-1)
\eeqa
where, as well known, the entropy of the solution is $S= \pi \sqrt{J_4}$. In the case at hand 
$\phi=0\,{\mbox{mod}}\,2\pi$ and the last term in the above equation drops out (according to the 
fact that it is a four, rather than a five parameters solution). The above expression reduces to:
\beqa
J_4= s_0s_1s_2s_3
\eeqa
where, using relations (\ref{chargeqp}), it follows:
\beqa
s_0&\equiv&(|Z_1|+|Z_2|+|Z_3|+|Z_4|)=\sqrt{2}q_0 \nn \\
s_1&\equiv&(|Z_1|-|Z_2|-|Z_3|+|Z_4|)=\sqrt{2}p^1 \nn \\
s_2&\equiv&(-|Z_1|+|Z_2|-|Z_3|+|Z_4|)=\sqrt{2}p^2 \nn \\
s_3&\equiv&(-|Z_1|-|Z_2|+|Z_3|+|Z_4|)=\sqrt{2}p^3 
\eeqa
As noticed in \cite{malser}, the charge vector basis we have chosen turns out to be the suitable 
one for the microscopic identification, as for reading off the values of the integers $N_I$ 
from the relations (\ref{chargeqp}). First notice that the $4$ dimensional charge of a wrapped 
D$p$--brane is $Q_p\,=\,\hat \mu_p \cdot V_p/\sqrt{V_6}$ where 
$\hat \mu_p\,=\,\sqrt{2\pi}(2\pi\sqrt{\alpha'})^{3-p}$ is the normalized D$p$--brane charge 
density in ten dimensions. Provided the asymptotic values of the dilatons, which parameterize 
the radii of the compactifying torus and which has been taken to be unitary, it turns out that, 
in units where $\alpha'=1$, the four dimensional quanta of charge for {\it any} kind of (wrapped) 
D$p$--brane is equal to 
$\sqrt{2\pi}$. On the contrary, our quantized charges $(p^\Lambda,q_\Lambda)$ are integer valued. 
The entropy formula (\ref{entrqp}) is reproduced microscopically by the above D--branes 
configuration, table \ref{NDbrane}, if we have precisely 
$N_0=q_0\,,\,N_1=p^1\,,\,N_2=p^2\,,\,N_3=p^3$, consistently with the result of our previous geometrical analysis. Indeed the microscopic entropy counting for this 
configuration has been performed in \cite{bala1} and gives:
\beqa
\label{entrmicr}
S_{micro}\,=\,2\pi\sqrt{N_0N_1N_2N_3}
\eeqa
which exactly matches expression (\ref{entrqp}). From the configuration in table \ref{NDbrane} 
one can obtain, by various dualities, other four parameters solutions. For instance, $T$--dualizing 
on the whole $T^6$, one obtains a configuration made of $N_0$ D$6$--branes and 3 bunches of 
($N_1,N_2,N_3$) D$2$--branes on the planes ($x^4,x^5$), ($x^6,x^7$), ($x^8,x^9$) 
respectively.

\section{Discussion}

The main aim of the present article was to define in a precise
mathematical fashion a connection between the {\it macroscopic}
analysis of $1/8$ BPS black hole solutions of $N=8,\, d=4$
supergravity carried out in \cite{mp1,mp2,bft,bft2} and the {\it
microscopic} description of the subclass of these solutions carrying
R--R charge in terms of D--branes.  To this end it was necessary to
single out in the $U$--duality orbit of $1/8$ BPS black holes those
charged with respect to R--R vector fields, once the $N=8$ theory is
interpreted as the low--energy limit of type II superstring on
$T^6$. The first step in this direction was to describe group
theoretically the embedding of a class of $STU$ models yielding the
generating solution of R--R charged $1/8$ BPS black holes within the
$d=4$ maximal supergravity. To achieve this we used the SLA techniques
developed in \cite{solv,RR} in order to characterize geometrically the
R--R and the NS--NS ten dimensional origin of the fields in the
$N=8,\, d=4$ theory. We improved them in order to provide also two  
distinct descriptions of  the
$N=8,\,d=4$ theory as deriving from type IIA or type IIB theories in
ten dimensions. A SLA characterization of the effect of $T$--duality
transformations on compact directions is also formulated in terms of
the action of  automorphisms of the $S\times T$--duality group
$SL(2,\IR)\times SO(6,6)$ on the SLA generating the scalar manifold of
the theory. Besides characterizing the class of embeddings of the $STU$
model describing R--R charged generating solutions, we defined a $U$--dual 
class of $STU$ models describing NS--NS charged solutions in
the same mathematical fashion. The $U$--duality relation between
the two classes of $STU$ models can be inferred from their embedding
in the larger $N=8$ theory (this transformation is in the automorphism
group of $U=E_{7(7)}$ but not in $Aut(S\times T)=Aut(SL(2,\IR)\times
SO(6,6))$, since it does not preserve the R--R and NS--NS identities
of the fields, while the models within each class are related by
transformations in $Aut(S\times T)$ ). 

Eventually we focused on
two particular representatives of the R--R class of $STU$ models and
interpreted their fields in terms of  ten dimensional type IIA or type
IIB fields. As an example  a particular $4$--parameter solution of
this model was considered and a microscopic interpretation of it was
given in terms D--branes (a configuration of $D4$ and $D0$--branes, in
the framework of type IIA theory). 

The utility of the
mathematical apparatus constructed in the present article (whose
details can be found in the appendix) is more general. Indeed it
allows to characterize the bosonic sector of whatever model embedded
in the $N=8,\, d=4$ theory in terms of dimensionally reduced fields of
type IIA or IIB theories, and therefore to interpret microscopically
 its  solutions. Moreover, a precise prescription
is given as to how to act on a particular truncation of the
$N=8,\,d=4$ theory by means of $S\times T$ duality (at the classical
level). This method can be easily extended to describe also $U$--dual
embeddings by considering the action of the whole $Aut(E_{7(7)})$ 
on the  model. \par
Within the two $T$--dual R--R
embeddings of the $STU$ model discussed in subsection 2.2 it would be
interesting to find the generating solution in a form which is  easily
interpreted in terms of bound states of D--branes, recovering as an
example the two $T$--dual microscopic configurations described in
\cite{bala2}. One of these configurations consists of a system of D4
and D0--branes in type IIA theory analogous to the one represented in
table \ref{NDbrane} but with a magnetic  flux switched on
the world volume of one bunch of parallel D4--branes (which would imply
additional effective D2 and D0 charges,\cite{doug}).  As already
pointed out in the introduction, this flux would correspond in the
solution to the presence of non--trivial axions coming from NS--NS $B$ field 
components (which would contribute to a non vanishing real
part of the central charges). An embedding where to look for this
solution is the type IIA embedding described in subsection 2.2. The
type IIB configuration described in \cite{bala2} consists of 4 sets
of D3--branes at angles and is obtained from the one described above by
a $T$--duality along the internal directions $x^{5,7,9}$ (these angles
being related to the flux in the dual configuration and consistent
with the supersymmetry requirement,\cite{jab}).  The corresponding
solution has to be looked for in the type IIB embedding of section
2.2. In the limits in which the magnetic flux is sent to zero
or the system of D3--branes is set to be orthogonal \cite{behr}, one
recovers the four parameter solution described in the previous section
(or its $T$--dual).

The importance of finding both a  macroscopic and
miscoscopic description of the generating solution relies in the fact
that it would allow  to have a precise control on both the macroscopic
and microscopic structures of all regular stringy black holes related
by $U$--duality transformations, and hence sharing the same
entropy. Starting from a configuration for which a microscopic entropy
counting is known (as for instance pure D--branes configurations) one
can then have an entropy prediction and a description, both at
macroscopic and microscopic levels, of those configurations for which
a microscopic entropy counting is out of present reach (as for example
pure NS--NS state configurations). This could help  in revealing the
underlying common properties of  very different black holes sharing
the same entropy and hence giving some insights  into the basic
properties of stringy oriented microscopic entropy counting. 

Although there exist in the literature some macroscopic 5 parameter generating
solutions,  \cite{cve1,cve2,bft2}, the interpretation, at the
microscopic level, of the  parameters entering these solutions is
quite difficult, especially as far as the  fifth one is
concerned. Hence a simple and clear description of the generating
solution,  both at macroscopic and microscopic level, is still
missing. The completion of this  program is left to a future work.

\vskip 10pt

{\bf Acknowledgments}

We would like to thank V. Balasubramanian and F. Larsen for very
useful discussions and the organizers of the ICTP conference on
``Black holes Physics'' during which part of this work was done. We
also acknowledge useful discussions with A. Hammou, F. Morales,
C.A. Scrucca, R. Russo, M.  Serone, G. Bonelli, L. Andrianopoli,
R. D'Auria and S. Ferrara. Finally, we  would like to thank P. Fr\`e
for reading a preliminary version of the manuscript.  One of us, M.T.,
would like to thank SISSA for the kind hospitality.

\appendix
\section*{Appendix A: SLA Description of $\mathbf {N=8,\,d=4}$ from 
Type IIA and type IIB Theories and $\mathbf {S,T}$--dualities.}
\setcounter{equation}{0}
\addtocounter{section}{1}

{\bf \underline{ SLA of ${\cal E}_{7(7)}^\pm$ and the scalar
fields:}}\\
As already stated in section 2, in the SLA representation of a theory,
the scalar fields are parameters of the solvable Lie algebra
generating the scalar manifold (in most supergravities at the
classical level the scalar manifold can be described as a solvable
group manifold). The $N=8$ theory in four dimensions can be
interpreted as the low energy limit of type IIA or type IIB  theories
on $T^6$. Depending on the two interpretations we shall give two
different SLA descriptions of the scalar manifold which are consistent
with the geometric characterization of $T$--duality to be given in the
sequel.

As anticipated in subsection 2.1 one may construct the ${\cal
E}_{7(7)}$ algebra in two ways, depending on whether the $D_6$ Dynkin
diagram consisting of the roots $(\alpha_i)_{i=1,\dots,6}$ in
(\ref{dynke7pm}) is extended by attaching the highest weight of the
${\bf 32}^+$ of $SO(6,6)$ ($\alpha_7^+$) to $\alpha_5$ or the  highest
weight of the ${\bf 32}^-$ ($\alpha_7^-$) to $\alpha_6$. Thus we
obtain two   ${\cal E}_{7(7)}$ isomorphic algebras, namely  ${\cal
E}_{7(7)}^\pm$. As previously stated we parameterize with the scalars
of the $N=8$ theory deriving from a dimensional reduction of type IIA
of type IIB theory the SLA of ${\cal E}_{7(7)}^+$ and ${\cal
E}_{7(7)}^-$  respectively.

The SLA is generated by the non--compact Cartan generators (in the case of maximal supergravities
 the whole Cartan subalgebra contributes to the SLA) and the shift
operators corresponding to roots with positive restriction on the
non--compact Cartan generators.\footnote{The criterion of positivity
is defined for instance by fixing an orthonormal basis of non--compact
Cartan generators with a certain order and ordering the restrictions
of the roots to this basis in a lexicographic way: e.g. with respect
to the system $H_{\epsilon_1},H_{\epsilon_2}$ the root
$\epsilon_1-\epsilon_2$ is positive while with respect to
$H_{\epsilon_2},H_{\epsilon_1}$ the root $\epsilon_2-\epsilon_1$ is
positive. }  As far as the common NS--NS sector is concerned, a
suitable basis of non--compact Cartan generators will be parametrized
by the radii of the torus and by the ten dimensional dilaton:
\begin{eqnarray}
{\cal C}_K ({\rm IIB})&=&\sum_{i=1}^6-\sigma_i
H_{\epsilon_i+\frac{\epsilon_7} {\sqrt{2}}} +\phi
H_{\sqrt{2}\epsilon_7}\nonumber\\ {\cal C}_K ({\rm
IIA})&=&\sum_{i=1}^5-\sigma_i H_{\epsilon_i+\frac{\epsilon_7}
{\sqrt{2}}} -\sigma_6 H_{-\epsilon_6+\frac{\epsilon_7}{\sqrt{2}}}+
\phi H_{\sqrt{2}\epsilon_7}\nonumber\\ \sigma_i&=& {\rm
ln}(\rho_{i+3})
\label{strange}
\end{eqnarray}
where, as previously stated, $\rho_k$ ($k=4,\dots ,9$) are the radii of the internal torus 
along the directions $x^k$.\footnote{They may be characterized in terms of the diagonal elements 
of the vielbein matrix $V^{\hat{k}}_k$ of the torus 
in a basis in which it is triangular and the square of their product gives 
therefore the determinant of the metric $G_{ij}$.} 
In the expressions (\ref{strange})  the overall coefficient of
$H_{\sqrt{2}\epsilon_7}$  is the four dimensional dilaton:
\begin{eqnarray}
\phi_4&=& \phi-\frac{1}{2}\sum_{i=1}^6\sigma_i=\phi-\frac{1}{4}
{\rm ln}\left({\rm det}(G_{ij})\right)
\end{eqnarray}
The strange non--orthonormal basis in (\ref{strange}) is defined by
the decomposition of  the $U$--duality group in $d$ dimensions with
respect to the $U$--duality group in $d+1$  dimensions for maximal
supergravities \cite{solv}:
\begin{eqnarray}
{\cal E}_{r+1(r+1)}&\rightarrow & O(1,1)_r+{\cal
E}_{r(r)}\,\,\,(r=10-d)
\end{eqnarray} 
where ${\cal E}_{r(r)}$ is obtained by deleting the extreme root in
the Dynkin diagram  of ${\cal E}_{r+1(r+1)}$ (on the branch of
$\alpha_1,\,\alpha_2,\dots$) and substituting  it with the Cartan
generator $O(1,1)_r$ orthogonal to the rest of the Dynkin diagram.
These $O(1,1)_r$ define the basis in (\ref{strange}) and are naturally
parametrized by  $-\sigma_{7-r}$.

As far as the the remaining NS--NS fields are concerned they
parameterize the positive roots of $SL(2,\IR)\times
SO(6,6)$. According to the definition of the ordering relation among
the roots with respect to the orthonormal basis $(H_{\epsilon_i})$,
which determines the roots contributing to the SLA, one can have
different equivalent SLAs, usually related by automorphisms of
$D_6$. It is natural to associate always the fields $G_{ij}$ ($i\neq
j$) with the roots $\pm (\epsilon_i-\epsilon_j)$ and  $B_{ij}$ with
the roots $\pm (\epsilon_i+\epsilon_j)$, since in the representation
${\bf 12}$ of $SO(6,6)$ in which the generators have the form:
\begin{eqnarray}
\left(M_{\Lambda\Sigma}\right)_\Delta^{\phantom{\delta}\Gamma}&=&\eta_{\Lambda\Delta}
\delta_\Sigma^\Gamma-\eta_{\Sigma\Delta}\delta_\Lambda^\Gamma\nonumber\\
\eta_{\Lambda\Delta}&=& {\rm diag}(++++++------)\nonumber\\
H_{\epsilon_i}&=& M_{ii+6}
\end{eqnarray}  
the shift operators corresponding to the former roots have a symmetric
$6\times 6$ off--diagonal block, while those corresponding to the
latter roots have an antisymmetric off--diagonal block. Using a
lexicographic ordering with respect to  the basis $(H_{\epsilon_i})$
the roots contributing to the SLA and the corresponding scalar fields
are listed in table 2.

Finally the R--R fields, as already pointed out, parameterize the
shift operators corresponding to the roots which are weights of the
${\bf 32}^\pm$ of $SO(6,6)$. These roots are:
\begin{eqnarray}
{\bf 32}^+ :\,\,\alpha^+ &=&-\frac{1}{2}\left(\pm \epsilon_1\pm
\epsilon_2\pm \epsilon_3\pm \epsilon_4\pm \epsilon_5\pm
\epsilon_6\right)+\frac{\sqrt{2}}{2}\epsilon_7\nonumber\\ &&
(\mbox{odd number of ``$+$'' signs within brackets})\nonumber\\ {\bf
32}^- :\,\,\alpha^- &=&-\frac{1}{2}\left(\pm \epsilon_1\pm
\epsilon_2\pm \epsilon_3\pm \epsilon_4\pm \epsilon_5\pm
\epsilon_6\right)+\frac{\sqrt{2}}{2}\epsilon_7\nonumber\\
&&(\mbox{even number of ``$+$'' signs within brackets})
\end{eqnarray}
Indeed the {\it chirality} operator $\gamma$ is easily computed in
terms of the product of the Cartan  generators
$(H_{\epsilon_i})_{i=1,\dots,6}$ in the spinorial representation
($(S(H_{\epsilon_i}))_{i=1,\dots,6}$):
\begin{eqnarray}
\{\gamma_\Lambda,\gamma_\Sigma\}&=&2 \eta_{\Lambda\Sigma}\nonumber\\
S(H_{\epsilon_i}) &\equiv &
\gamma_i\gamma_{i+6}\,,\,\,(i,1,\dots,6)\nonumber\\ \gamma &=&
\gamma_1\gamma_2\cdots \gamma_{12}=-
S(H_{\epsilon_1})S(H_{\epsilon_2})\cdots S(H_{\epsilon_6})
\end{eqnarray}
it is easy to check that $\gamma$ is positive on the $\alpha^+$ and
negative on the $\alpha^-$. A precise correspondence between the
spinorial roots and scalar fields from type IIA and type IIB theories
is again given in table 2.

\noindent
{\bf \underline {S, T--duality:}}\\    
 The aim of the following discussion is to  characterize the effect (at
the classical level) of $S$ and $T$--duality on the embedding of a theory 
in the $N=8$ one, from the point of view of the scalar fields, as the action 
of transformations in $Aut(S\times T)$ on the SLA generating the scalar
manifold.\par
 Automorphisms of a semisimple Lie algebra G are isomorphisms
of the algebra on  itself and can be {\it inner} if their action can
be expressed as a conjugation of the algebra by means of a group
element generated by the algebra itself, or {\it outer} if they do not
admit such a representation (see for instance \cite{jacob}). A generic automorphism may be reduced, through 
the composition with a suitable (nilpotent) inner automorphism, 
to an isometric mapping which leaves the Cartan subalgebra stable.
Let us focus on the latter kind of transformations, which we shall denote by $\psi_\tau$. It can be shown that the restriction of the group $\{\psi_\tau\}$
 to the Cartan subalgebra is isomorphic to the automorphism group 
of the root space $\Delta$ consisting of the transformations $\tau$ on the weight lattice leaving the Cartan--Killing matrix invariant ({\it rotations}).
It can be shown that inner automorphisms $\psi_\tau$ correspond to $\tau$
in the Weyl group of G while in the case of outer $\psi_\tau$, $\tau$ may be reduced, modulo Weyl transformations, to symmetries of the Dynkin diagram (permutations of the simple roots).\par
Conversely, given a rotation $\tau$ on $\Delta$, one may associate with it an
automorphism $\tilde{\psi}_\tau$ on the whole G whose action on its canonical basis reads:
\begin{eqnarray}
\tilde{\psi}_\tau(H_\beta)&=&H_{\tau(\beta)}\,\,;\,\,\tilde{\psi}_\tau(E_\alpha) \propto E_{\tau(\alpha)}\nonumber\\
\alpha,\beta \,&&\mbox{roots}
\label{simplerecipe}
\end{eqnarray}
A general $\psi_\tau$ has the form $\psi_\tau=\tilde{\psi}_\tau \cdot \omega$,
where $\omega$ is an automorphism leaving the Cartan subalgebra of G pointwise fixed (these automorphisms are all inner \cite{jacob}).\par
In the case  in which $G=SL(2,\IR)\times SO(6,6)$ the rotation corresponding to an outer
automorphism $\psi_\tau$ can be reduced (modulo Weyl transformations) 
to the only
symmetry transformation of $D_6$, i.e. $\alpha_5\leftrightarrow
\alpha_6$, or equivalently $\epsilon_6\rightarrow -\epsilon_6$. It can
be shown in particular that rotations on the root space
amounting to a change of sign of an odd number of $\epsilon_i$ ($i=1,\dots, 6$)
 define outer automorphisms. Since the automorphisms preserve algebraic
structures, they will map  solvable subalgebras into solvable
subalgebras. Of course we do not expect  all the automorphisms of $S\times T$
to be automorphisms of ${\cal E}_{7(7)}$ since, for instance,
 the Dynkin diagram of
the latter does not have symmetries. Indeed it is easy to check that
outer automorphisms of $SO(6,6)$ map ${\cal E}_{7(7)}^\pm$ into ${\cal
E}_{7(7)}^\mp$ (this derives from the fact that  changing sign to an
odd number of $\epsilon_i$ maps $\alpha^\pm$ into $\alpha^\mp$).

{\it We characterize algebraically a T--duality transformation ``(large radius)  $\leftrightarrow $  (small radius)'' along a
compact direction $x^k$ ($k=4,\dots,9$) as the action of an outer
automorphism $\psi_\tau$ corresponding to 
$\tau:\,\,\epsilon_{k-3}\rightarrow -\epsilon_{k-3}$, and an
S--duality transformation  ``(strong coupling) $\leftrightarrow $ (weak coupling)'' as a $\psi_\tau$ such that $\tau:\,\,\epsilon_7\rightarrow
-\epsilon_7$.}

Let us consider as an example a $T$--duality transformation along the
direction  $x^9$ and its effects on $\rho_9$ ($\sigma_6$) and
the dilaton $\phi$ starting from type IIB fields. According to the
geometrical prescription given  above:
\begin{eqnarray}                       
\left(\phi-\frac{\sigma_6}{2}\right) H_{\sqrt{2}\epsilon_{7}}-\sigma_6
H_{\epsilon_6}&=&  \left(\phi^\prime-\frac{\sigma^\prime_6}{2}\right)
H_{\sqrt{2}\epsilon_{7}}- \sigma^\prime_6 H_{-\epsilon_6}\nonumber\\
\sigma^\prime_6 =-\sigma_6\,&\Rightarrow &
\rho_9^\prime=1/\rho_9\nonumber\\ \phi^\prime = \phi-\sigma_6
&=&\phi-{\rm ln}(\rho_9)
\label{tdualphi}
\end{eqnarray}
where the primed fields are the corresponding type IIA fields and the
last equation  is the known transformation rule for the dilaton under
$T$--duality along a compact  direction (in the units
$\alpha^\prime=1$).  As far as the other fields are concerned, the
action of this automorphism is to map the roots
$\epsilon_i\pm\epsilon_6$ into $\epsilon_i\mp\epsilon_6$. If we extend 
the rotation $\tau:\,\,\epsilon_6\rightarrow -\epsilon_6$ to the whole Lie algebra using the simple recipe (\ref{simplerecipe}) the fields $G_{i9}$ and 
$B_{i9}$ will be mapped (modulo proportionality constants $c_{1,2}$ 
to be fixed) into $B^\prime_{i9}$ and $G^\prime_{i9}$ respectively:
\begin{eqnarray}
G_{i9}E_{\epsilon_i-\epsilon_9}+B_{i9}E_{\epsilon_i+\epsilon_9}=
G_{i9}E_{\epsilon_i+(-\epsilon_9)}+B_{i9}E_{\epsilon_i-(-\epsilon_9)}=
c_1 B^\prime_{i9}E_{\epsilon_i+(-\epsilon_9)}+c_2 G^\prime_{i9}E_{\epsilon_i-(-\epsilon_9)} \nn \\
\end{eqnarray}
The transformation on the R--R fields, applying a similar rationale,
can be read off table 2.

{\bf \underline{Vector fields:}}\\  As far as the vector fields are
concerned, we refer to the conventions of \cite{mp1} in which the
corresponding representation $Sp(56)_D$
\footnote{By $Sp(56)_D$ we denote the ${\bf 56}$ symplectic
representation of $E_{7(7)}$ in which the Cartan generators are
diagonal.} of ${\cal E}_{7(7)}$ was described in terms of 56 weights
$W^{(\lambda)}$ ($\lambda=1,\dots,56$), whose difference from the {\it
highest weight} $W^{(51)}$ are suitable combinations of the simple
roots with positive integer coefficients (the first 28 weights
correspond to {\it magnetic} charges, the last 28 to {\it electric}
charges). In the conventions introduced in this paper, depending on
whether we consider ${\cal E}_{7(7)}^{\pm}$ (type IIA/IIB) we have two
set of weights ${W^{(\lambda)}}^{\mp}$. These weights provide a suitable
basis also for the two  representations ${\bf 28}$ and ${\bf
\bar{28}}$ in which the ${\bf 56}$ decomposes  with respect to
$SU(8)$: the  ${\bf 28}$ is generated by $W^{(\alpha)}$ with
$\alpha=1,\dots,28$ and the  ${\bf \bar{28}}$ by
$W^{(\alpha+28)}=-W^{(\alpha)}$.  We talk about representation ${\bf
56}$ when we consider the quantized charges  $(p^\alpha,q_\beta)$ and
of representation ${\bf 28}+{\bf \bar{28}}$ referring to  the {\it
dressed} charges to be defined below. The weights $W^{(\lambda)}$  can
be naturally put in correspondence with the vector fields obtained
from the dimensional reduction of the  type IIA or IIB theory
respectively. Both the first 28 magnetic charges and the last 28
electric charges decompose into a first set of 16 R--R charges (which
contribute to a $ {\bf 32}^\pm $ of $SO(6,6)$) and a second set of 12
NS--NS charges. Representing these weights (as well as the roots for
the scalar fields in table 2) in the basis of
$(\epsilon_i)_{i=1,\dots,7}$ the correspondence weights
$\leftrightarrow $ vectors (or roots $\leftrightarrow $ scalars)
becomes natural and consistent with our characterization of $S\times T$
duality. Indeed, as far as the R--R fields are concerned, the natural
correspondence is between the inner indices of the dimensionally
reduced form (which gives rise either to a scalar or to a vector) and
the number and positions of the ``$+$'' signs multiplying the
$(\epsilon_i)_{i=1,\dots,6}$ in the corresponding weight.\footnote{For
example the vector $A_{\mu ijkl}$ corresponds to the weight $(1/2)
(..+_i..+_j..+_k..+_l..)$.}  In tables 2 and 3 this correspondence has
been ``nailed'' down for a particular  $S\times T$--duality gauge (so
that the fields (weights) of IIB and  IIA are related by a
$T$--duality along the compact direction $x^9$ (automorphism
$\epsilon_6\rightarrow -\epsilon_6$)), making it possible to infer the
transformation rules  of the fields under a generic $S\times T$
transformation.

We would like ultimately to be able to use table 3
in order to infer which  kind of D--brane charges characterize a
particular solution and from it infer the particular  supersymmetric
brane configuration which reproduces it (just as we have done for the
relatively simple example of the four parameter solution).  Table 3
has to be applied to the vector in the ${\bf 28}+{\bf \bar{28}}$ of
$SU(8)$  consisting of the following {\it dressed} charges (i.e. the
charges of the branes coupled  to the moduli of the internal torus
\footnote{In the general case also to the R--R scalars.}, which are
the effective microscopic physical charges one could measure):
\begin{eqnarray}
\left(\matrix{y^\alpha (\phi)\cr x_\beta
(\phi)}\right)&=&\IC\,\IL^t(\phi)\, \IC  \left(\matrix{p^\alpha\cr
q_\beta}\right)=-\IL^{-1}(\phi)\left(\matrix{p^\alpha\cr
q_\beta}\right)
\label{interpret}
\end{eqnarray}
where we have used the notation of \cite{bft}\footnote{If the vector
 on the right hand side of (\ref{interpret}) was projected on the
 Young basis of the ${\bf 28}+{\bf \bar{28}}$ of $SU(8)$ we would have
 obtained on the left hand side the central charge vector
 $(Z^{AB},\bar{Z}_{AB})$ \cite{mp1}.}: $\phi$ is the point of the
 moduli space at radial infinity, $\IL(\phi)$ is the coset
 representative of the scalar manifold (solvable group element) in its
 $Sp(56)_D$ representation, $\IC$ is the symplectic invariant matrix
 and $(p^\alpha,q_\beta)$ are the usual quantized charges
 ($\alpha,\beta=1,\dots, 28$). \par If we consider, as we did in the
 present paper, a theory embedded in the $N=8,\,d=4$ one (i.e. a
 consistent truncation), then the embedding can be {\it perturbative}
 or {\it non--perturbative} depending on whether the vectors described
 in the Lagrangian of the truncation derive from the same forms as the
 $28$ vectors appearing in the Lagrangian of the larger theory, or
 some of them derive from the dimensional reduction of the
 corresponding Hodge dual forms. From the geometrical point of view
 these two situations are represented respectively by the case in
 which the $y^n$ ($n=1,\dots ,n_v\,(\#\mbox{ of vectors})$) span a
 space generated by a subset of the {\it magnetic}  weights
 $W^{(\alpha)}$ only (and therefore the $x_n$ are expressed in the
 basis of the {\it electric} weights $W^{(\alpha+28)}$), or by the
 case in which $y^n$ is also expressed in terms of part of the {\it
 electric} weights $W^{(\alpha+28)}$.  The correspondence between the electric
 and magnetic charges $(y^n,x_n)$  and the weights $W^{(\lambda)}$ may
 be inferred from the embedding of the SLA which generates the  scalar
 manifold of the truncation, $Solv_{trunc}$, within the  $Solv({\cal
 E}_{7(7)}^\pm)$ of the larger theory (which defines which scalar
 fields are switched off in the process of truncation). Indeed, once
 the  Cartan generators $H_{\gamma_i}$ in $Solv_{trunc}$ are
 expressed in the $Sp(2 n_v)_D$ representation in which   they are
 diagonal, interpreting their diagonal elements as the scalar product
 of the vectors $\gamma_i$ and $2 n_v$ of the $W^{(\lambda)}s$ it is
 possible to  characterize the embedding consistently in terms of
 vector fields as well.\par This procedure was applied in section 3
 for the type IIA embedding of the $STU$ model described in subsection
 2.2. From the form of the Cartan generators in the $Sp(8)_D$ we could
 give a consistent interpretation of the electric and magnetic charges
 ($x_n$ and $y^n$) in terms of the weights $W^{(\lambda)}$
 (eqs. (\ref{iden})) and thus, applying table 3, we were able to tell
 which of the ten dimensional forms these charges were associated
 with. In the dilatonic solution discussed in section 3, the electric
 charge $x_0=-q_0$, corresponds to the weight $W^{(1)}$ which, in the
 original $N=8$ theory is magnetic and therefore the embedding is
 non--perturbative. The 3 magnetic charges $y^i$ coincide with $-p^i$
 which in turn correspond to magnetic weights of the $N=8$ theory (the
 one to one correspondence between the dressed charges $(y,x)$ and the
 quantized ones $(p,q)$ on our solution is related to the fact that
 the solution is dilatonic and that we are considering a square torus
 ($b_i=-1$)). If axions were switched on the correspondence between
 dressed and quantized charges would be less trivial (since the coset
 representative would not be diagonal) and table 3 could give the
 microscopic interpretation of the dressed charges only. \par  In the
 literature the vector of dressed charges $(y(\phi),x(\phi))$ in the
 $N=8$ theory is also written in the following equivalent complex form:
\begin{eqnarray}
x_{ij} (\phi)+{\rm i} y_{ij} (\phi)&=&-\frac{1}{\sqrt{2}}\left(\Gamma^{AB}\right)_{ij}Z_{AB}
\end{eqnarray}
where $\left(\Gamma^{AB}\right)_{ij}$ is an $SO(8)$ rotation and the antisymmetric couple 
of indices $(ij)$ run in the ${\bf 28}$ of $SO(8)$.

\begin{table} [ht]
\vskip -25pt
\begin{center}
{\tiny
\begin{tabular}{|c|c|c|c|c|}
\hline
{\small IIA} & {\small IIB} & {\small $\alpha_{m,n}^\mp$ ($IIB/IIA$) } & {\small $\epsilon_i$--components} 
& {\small $\alpha_i$--components} \\
\hline
\hline
$A_{9}$ & $\rho$ & $\alpha_{1,1}$ & $\frac{1}{2}(-1,-1,-1,-1,-1, {\mp} 1,\sqrt{2})$ & $(0,0,0,0,0,0,1)$\\
\hline
\hline
$B_{8\,9}$ & $B_{8\,9}$ & $\alpha_{2,1}$ & $(0,0,0,0,1,1,0)$ & $(0,0,0,0,0,1,0)$\\
\hline
$G_{8\,9}$ & $G_{8\,9}$ & $\alpha_{2,2}$ & $(0,0,0,0,1,-1,0)$ & $(0,0,0,0,1,0,0)$\\
\hline
$A_{8}$ & $A_{8\, 9}$ & $\alpha_{2,3}$ & $\frac{1}{2}(-1,-1,-1,-1,1, {\pm} 1,\sqrt{2})$ & 
$(0,0,0,0,0,1,1)$\\
\hline
\hline
$B_{7\,8}$ & $B_{7\,8}$ & $\alpha_{3,1}$ & $(0,0,0,1,1,0,0)$ & $(0,0,0,1,1,1,0)$\\
\hline
$G_{7\,8}$ & $G_{7\,8}$ & $\alpha_{3,2}$ & $(0,0,0,1,-1,0,0)$ & $(0,0,0,1,0,0,0)$\\
\hline
$B_{7\,9}$ & $B_{7\,9}$ & $\alpha_{3,3}$ & $(0,0,0,1,0,1,0)$ & $(0,0,0,1,0,1,0)$\\
\hline
$G_{7\,9}$ & $G_{7\,9}$ & $\alpha_{3,4}$ & $(0,0,0,1,0,-1,0)$ & $(0,0,0,1,1,0,0)$\\
\hline
$A_{7\,8\,9}$ & $A_{7\, 8}$ & $\alpha_{3,4}$ & $\frac{1}{2}(-1,-1,-1,1,1, {\mp} 1,\sqrt{2})$ & 
$(0,0,0,1,1,1,1)$\\
\hline
$A_{7}$ & $A_{7\, 9}$ & $\alpha_{3,6}$ & $\frac{1}{2}(-1,-1,-1,1,-1,{\pm} 1,\sqrt{2})$ & 
$(0,0,0,1,0,1,1)$\\
\hline
\hline
$B_{6\,7}$ & $B_{6\,7}$ & $\alpha_{4,1}$ & $(0,0,1,1,0,0,0)$ & $(0,0,1,2,1,1,0)$\\
\hline
$G_{6\,7}$ & $G_{6\,7}$ & $\alpha_{4,2}$ & $(0,0,1,-1,0,0,0)$ & $(0,0,1,0,0,0,0)$\\
\hline
$B_{6\,8}$ & $B_{6\,8}$ & $\alpha_{4,3}$ & $(0,0,1,0,1,0,0)$ & $(0,0,1,1,1,1,0)$\\
\hline
$G_{6\,8}$ & $G_{6\,8}$ & $\alpha_{4,4}$ & $(0,0,1,0,-1,0,0)$ & $(0,0,1,1,0,0,0)$\\
\hline
$B_{6\,9}$ & $B_{6\,9}$ & $\alpha_{4,4}$ & $(0,0,1,0,0,1,0)$ & $(0,0,1,1,0,1,0)$\\
\hline
$G_{6\,9}$ & $G_{6\,9}$ & $\alpha_{4,6}$ & $(0,0,1,0,0,-1,0)$ & $(0,0,1,1,1,0,0)$\\
\hline
$A_{6\,7\,9}$ & $A_{6\, 7}$ & $\alpha_{4,7}$ & $\frac{1}{2}(-1,-1,1,1,-1, {\mp} 1,\sqrt{2})$ & 
$(0,0,1,2,1,1,1)$\\
\hline
$A_{6\,8\,9}$ & $A_{6\, 8}$ & $\alpha_{4,8}$ & $\frac{1}{2}(-1,-1,1,-1,1,{\mp} 1,\sqrt{2})$ & 
$(0,0,1,1,1,1,1)$\\
\hline
$A_{6}$ & $A_{6\, 9}$ & $\alpha_{4,9}$ & $\frac{1}{2}(-1,-1,1,-1,-1,{\pm} 1,\sqrt{2})$ & 
$(0,0,1,1,0,1,1)$\\
\hline
$A_{6\,7\,8}$ & $A_{6\, 7\,8\,9}$ & $\alpha_{4,10}$ & $\frac{1}{2}(-1,-1,1,1,1,
 {\pm} 1,\sqrt{2})$ & 
$(0,0,1,2,1,2,1)$\\
\hline
\hline
$B_{5\,6}$ & $B_{5\,6}$ & $\alpha_{5,1}$ & $(0,1,1,0,0,0,0)$ & $(0,1,2,2,1,1,0)$\\
\hline
$G_{5\,6}$ & $G_{5\,6}$ & $\alpha_{5,2}$ & $(0,1,-1,0,0,0,0)$ & $(0,1,0,0,0,0,0)$\\
\hline
$B_{5\,7}$ & $B_{5\,7}$ & $\alpha_{5,3}$ & $(0,1,0,1,0,0,0)$ & $(0,1,1,2,1,1,0)$\\
\hline
$G_{5\,7}$ & $G_{5\,7}$ & $\alpha_{5,4}$ & $(0,1,0,-1,0,0,0)$ & $(0,0,1,0,0,0,0)$\\
\hline
$B_{5\,8}$ & $B_{5\,8}$ & $\alpha_{5,5}$ & $(0,1,0,0,1,0,0)$ & $(0,1,1,1,1,1,0)$\\
\hline
$G_{5\,8}$ & $G_{5\,8}$ & $\alpha_{5,6}$ & $(0,1,0,0,-1,0,0)$ & $(0,0,0,1,0,0,0)$\\
\hline
$B_{5\,9}$ & $B_{5\,9}$ & $\alpha_{5,7}$ & $(0,1,0,0,0,1,0)$ & $(0,1,1,1,0,1,0)$\\
\hline
$G_{5\,9}$ & $G_{5\,9}$ & $\alpha_{5,8}$ & $(0,1,0,0,0,-1,0)$ & $(0,0,0,0,1,0,0)$\\
\hline
$A_{5\,6\,9}$ & $A_{5\, 6}$ & $\alpha_{5,9}$ & $\frac{1}{2}(-1,1,1,-1,-1,{\mp} 1,\sqrt{2})$ & 
$(0,1,2,2,1,1,1)$\\
\hline
$A_{5\,7\,9}$ & $A_{5\, 7}$ & $\alpha_{5,10}$ & $\frac{1}{2}(-1,1,-1,1,-1, {\mp} 1,\sqrt{2})$ & 
$(0,1,1,2,1,1,1)$\\
\hline
$A_{5\,8\,9}$ & $A_{5\, 8}$ & $\alpha_{5,11}$ & $\frac{1}{2}(-1,1,-1,-1,1, {\mp} 1,\sqrt{2})$ & 
$(0,1,1,1,1,1,1)$\\
\hline
$A_{5}$ & $A_{5\, 9}$ & $\alpha_{5,12}$ & $\frac{1}{2}(-1,1,-1,-1,-1,{\pm} 1,\sqrt{2})$ & 
$(0,1,1,1,0,1,1)$\\
\hline
$A_{5\,7\,8}$ & $A_{5\, 7\,8\,9}$ & $\alpha_{5,13}$ & $\frac{1}{2}(-1,1,-1,1,1, {\pm} 1,\sqrt{2})$ & 
$(0,1,1,2,1,2,1)$\\
\hline
$A_{5\,6\,8}$ & $A_{5\,6\,8\, 9}$ & $\alpha_{5,14}$ & $\frac{1}{2}(-1,1,1,-1,1,
{\pm} 1,\sqrt{2})$ & 
$(0,1,2,2,1,2,1)$\\
\hline
$A_{5\,6\,7}$ & $A_{5\,6\,7\, 9}$ & $\alpha_{5,15}$ & $\frac{1}{2}(-1,1,1,1,-1,
{\pm} 1,\sqrt{2})$ & 
$(0,1,2,3,1,2,1)$\\
\hline
$A_{\mu\nu\rho}$ & $A_{5\,6\,7\, 8}$ & $\alpha_{5,16}$ & $\frac{1}{2}(-1,1,1,1,1,{\mp} 1,\sqrt{2})$ & 
$(0,1,2,3,1,2,1)$\\
\hline
\hline
$B_{4\,5}$ & $B_{4\,5}$ & $\alpha_{6,1}$ & $(1,1,0,0,0,0,0)$ & $(1,2,2,2,1,1,0)$\\
\hline
$G_{4\,5}$ & $G_{4\,5}$ & $\alpha_{6,2}$ & $(1,-1,0,0,0,0,0)$ & $(1,0,0,0,0,0,0)$\\
\hline
$B_{4\,6}$ & $B_{4\,6}$ & $\alpha_{6,3}$ & $(1,0,1,0,0,0,0)$ & $(1,1,2,2,1,1,0)$\\
\hline
$G_{4\,6}$ & $G_{4\,6}$ & $\alpha_{6,4}$ & $(1,0,-1,0,0,0,0)$ & $(1,1,0,0,0,0,0)$\\
\hline
$B_{4\,7}$ & $B_{4\,7}$ & $\alpha_{6,5}$ & $(1,0,0,1,0,0,0)$ & $(1,1,1,2,1,1,0)$\\
\hline
$G_{4\,7}$ & $G_{4\,7}$ & $\alpha_{6,6}$ & $(1,0,0,-1,0,0,0)$ & $(1,1,1,0,0,0,0)$\\
\hline
$B_{4\,8}$ & $B_{4\,8}$ & $\alpha_{6,7}$ & $(1,0,0,0,1,0,0)$ & $(1,1,1,1,1,1,0)$\\
\hline
$G_{4\,8}$ & $G_{4\,8}$ & $\alpha_{6,8}$ & $(1,0,0,0,-1,0,0)$ & $(1,1,1,1,0,0,0)$\\
\hline
$B_{4\,9}$ & $B_{4\,9}$ & $\alpha_{6,9}$ & $(1,0,0,0,0,1,0)$ & $(1,1,1,1,0,1,0)$\\
\hline
$G_{4\,9}$ & $G_{4\,9}$ & $\alpha_{6,10}$ & $(1,0,0,0,0,-1,0)$ & $(1,1,1,1,1,0,0)$\\
\hline
$B_{\mu\nu}$ & $B_{\mu\nu}$ & $\alpha_{6,11}$ & $(0,0,0,0,0,0,\sqrt{2})$ & 
$(1,2,3,4,2,3,2)$\\
\hline
$A_{\mu\nu\,9}$ & $A_{\mu\nu}$ & $\alpha_{6,12}$ & $\frac{1}{2}(1,1,1,1,1,{\pm} 1,\sqrt{2})$ & 
$(1,2,3,4,2,3,1)$\\
\hline
$A_{4\,5\,9}$ & $A_{4\,5}$ & $\alpha_{6,13}$ & $\frac{1}{2}(1,1,-1,-1,-1,{\mp} 1,\sqrt{2})$ & 
$(1,2,2,2,1,1,1)$\\
\hline
$A_{4\,6\,9}$ & $A_{4\,6}$ & $\alpha_{6,14}$ & $\frac{1}{2}(1,-1,1,-1,-1,{\mp} 1,\sqrt{2})$ & 
$(1,1,2,2,1,1,1)$\\
\hline
$A_{4\,7\,9}$ & $A_{4\,7}$ & $\alpha_{6,15}$ & $\frac{1}{2}(1,-1,-1,1,-1,{\mp} 1,\sqrt{2})$ & 
$(1,1,1,2,1,1,1)$\\
\hline
$A_{4\,8\,9}$ & $A_{4\,8}$ & $\alpha_{6,16}$ & $\frac{1}{2}(1,-1,-1,-1,1,{\mp} 1,\sqrt{2})$ & 
$(1,1,1,1,1,1,1)$\\
\hline
$A_{4}$ & $A_{4\,9}$ & $\alpha_{6,17}$ & $\frac{1}{2}(1,-1,-1,-1,-1,{\pm} 1,\sqrt{2})$ & 
$(1,1,1,1,0,1,1)$\\
\hline
$A_{4\,7\,8}$ & $A_{4\,7\,8\,9}$ & $\alpha_{6,18}$ & 
$\frac{1}{2}(1,-1,-1,1,1,{\pm} 1,\sqrt{2})$ & $(1,1,1,2,1,2,1)$\\
\hline
$A_{4\,6\,8}$ & $A_{4\,6\,8\,9}$ & $\alpha_{6,19}$ & 
$\frac{1}{2}(1,-1,1,-1,1,{\pm} 1,\sqrt{2})$ & $(1,1,2,2,1,2,1)$\\
\hline
$A_{4\,6\,7}$ & $A_{4\,6\,7\,9}$ & $\alpha_{6,20}$ & 
$\frac{1}{2}(1,-1,1,1,-1,{\pm} 1,\sqrt{2})$ & $(1,1,2,3,1,2,1)$\\
\hline
$A_{\mu\nu\,5}$ & $A_{4\,6\,7\,8}$ & $\alpha_{6,21}$ & 
$\frac{1}{2}(1,-1,1,1,1,{\mp} 1,\sqrt{2})$ & $(1,1,2,3,2,2,1)$\\
\hline
$A_{4\,5\,8}$ & $A_{4\,5\,8\,9}$ & $\alpha_{6,22}$ & 
$\frac{1}{2}(1,1,-1,-1,1,{\pm} 1,\sqrt{2})$ & $(1,2,2,2,1,2,1)$\\
\hline
$A_{4\,5\,7}$ & $A_{4\,5\,7\,9}$ & $\alpha_{6,23}$ & 
$\frac{1}{2}(1,1,-1,1,-1,{\pm} 1,\sqrt{2})$ & $(1,2,2,3,1,2,1)$\\
\hline
$A_{\mu\nu\,6}$ & $A_{4\,5\,7\,8}$ & $\alpha_{6,24}$ & 
$\frac{1}{2}(1,1,-1,1,1,{\mp} 1,\sqrt{2})$ & $(1,2,2,3,2,2,1)$\\
\hline
$A_{4\,5\,6}$ & $A_{4\,5\,6\,9}$ & $\alpha_{6,25}$ & 
$\frac{1}{2}(1,1,1,-1,-1,{\pm} 1,\sqrt{2})$ & $(1,2,3,3,1,2,1)$\\
\hline
$A_{\mu\nu\,7}$ & $A_{4\,5\,6\,8}$ & $\alpha_{6,26}$ & 
$\frac{1}{2}(1,1,1,-1,1,{\mp} 1,\sqrt{2})$ & $(1,2,3,3,2,2,1)$\\
\hline
$A_{\mu\nu\,8}$ & $A_{4\,5\,6\,7}$ & $\alpha_{6,27}$ & 
$\frac{1}{2}(1,1,1,1,-1,{\mp} 1,\sqrt{2})$ & $(1,2,3,4,2,2,1)$\\
\hline
\end{tabular}}
\end{center}
\caption{\small The correspondence between the positive 
roots $\alpha_{m,n}^\pm$ of the $U$--duality algebra ${\cal E}_{7(7)}^\pm$ 
and the scalar fields parameterizing the moduli space for either IIA or IIB 
compactifications on $T^6$. The notation $\alpha_{m,n}$ for the positive roots was 
introduced in \cite{solv}.}
\label{solfil}
\end{table}

\begin{table} [ht]
\vskip -45pt
\begin{center}
{\tiny
\begin{tabular}{|c|c|c|c|}
\hline
{\small IIA} & {\small IIB} & {\small ${W^{(\lambda)}}^{\pm}$ ($IIB/IIA$)} & {\small $\epsilon_i$--components: $IIB/IIA$} \\
\hline
\hline
$A_{\mu}$ & $A_{\mu 9}$ & $W^{(1)} $ & $\frac{1}{2}\left(-1,-1,-1,-1,-1,{\pm} 1,0 \right) $\\ \hline
$A_{\mu 5678}$ & $A_{\mu 56789}$ & $W^{(2)} $ & $\frac{1}{2}\left(-1,1,1,1,1,{\pm} 1,0 \right) $\\\hline
$A_{\mu 4678}$ & $A_{\mu 46789}$ & $W^{(3)} $ & $\frac{1}{2}\left(1,-1,1,1,1,{\pm} 1,0 \right) $\\\hline
$A_{\mu 4578}$ & $A_{\mu 45789}$ & $W^{(4)} $ & $\frac{1}{2}\left(1,1,-1,1,1,{\pm} 1,0 \right) $\\\hline
$A_{\mu 4568}$ & $A_{\mu 45689}$ & $W^{(5)} $ & $\frac{1}{2}\left( 1,1,1,-1,1,{\pm} 1,0\right) $\\\hline 
$A_{\mu 4567}$ & $A_{\mu 45679}$ & $W^{(6)} $ & $\frac{1}{2}\left(1,1,1,1,-1,{\pm} 1,0 \right) $\\\hline
$A_{\mu 6789}$ & $A_{\mu 678}$ & $W^{(7)} $ & $\frac{1}{2}\left(-1,-1,1,1,1,{\mp} 1,0 \right) $\\\hline
$A_{\mu 5789}$ & $A_{\mu 578}$ & $W^{(8)} $ & $\frac{1}{2}\left(-1,1,-1,1,1, {\mp} 1,0 \right) $\\\hline
$A_{\mu 5689}$ & $A_{\mu 568}$ & $W^{(9)} $ & $\frac{1}{2}\left( -1,1,1,-1,1,{\mp} 1,0 \right) $\\ \hline
$A_{\mu 5679}$ & $A_{\mu 567}$ & $W^{(10)} $ & $\frac{1}{2}\left(-1,1,1,1,-1,{\mp} 1,0 \right) $\\ \hline
$A_{\mu 4679}$ & $A_{\mu 467}$ & $W^{(11)} $ & $\frac{1}{2}\left(1,-1,1,1,-1,{\mp} 1,0 \right) $\\ \hline
$A_{\mu 4579}$ & $A_{\mu 457}$ & $W^{(12)} $ & $\frac{1}{2}\left(1,1,-1,1,-1,{\mp} 1,0 \right) $\\ \hline
$A_{\mu 4569}$ & $A_{\mu 456}$ & $W^{(13)} $ & $\frac{1}{2}\left( 1,1,1,-1,-1,{\mp} 1,0\right) $\\ \hline
$A_{\mu 4689}$ & $A_{\mu 468}$ & $W^{(14)} $ & $\frac{1}{2}\left(1,-1,1,-1,1,{\mp} 1,0 \right) $\\ \hline
$A_{\mu 4789}$ & $A_{\mu 478}$ & $W^{(15)} $ & $\frac{1}{2}\left(1,-1,-1,1,1,{\mp} 1,0 \right) $\\ \hline
$A_{\mu 4589}$ & $A_{\mu 458}$ & $W^{(16)} $ & $\frac{1}{2}\left(1,1,-1,-1,1,{\mp} 1,0 \right) $\\ \hline
$B_{\mu 4}$ & $B_{\mu 4}$ & $W^{(17)} $ & $\left(-1,0,0,0,0,0,{\frac{1}{{\sqrt{2}}}}\right)$\\ \hline
$B_{\mu 5}$ & $B_{\mu 5}$ & $W^{(18)} $ & $\left(0,-1,0,0,0,0,{\frac{1}{{\sqrt{2}}}}\right)$\\ \hline
$B_{\mu 6}$ & $B_{\mu 6}$ & $W^{(19)} $ & $\left(0,0,-1,0,0,0,{\frac{1}{{\sqrt{2}}}}\right)$\\ \hline
$B_{\mu 7}$ & $B_{\mu 7}$ & $W^{(20)} $ & $\left(0,0,0,-1,0,0,{\frac{1}{{\sqrt{2}}}}\right)$\\ \hline
$B_{\mu 8}$ & $B_{\mu 8}$ & $W^{(21)} $ & $\left(0,0,0,0,-1,0,{\frac{1}{{\sqrt{2}}}}\right)$\\ \hline
$G_{\mu 9}$ & $G_{\mu 9}$ & $W^{(22)} $ & $\left(0,0,0,0,0,1,{\frac{1}{{\sqrt{2}}}}\right)$\\ \hline
$G_{\mu 4}$ & $G_{\mu 4}$ & $W^{(23)} $ & $\left(-1,0,0,0,0,0,-{\frac{1}{{\sqrt{2}}}}\right)$\\ \hline
$G_{\mu 5}$ & $G_{\mu 5}$ & $W^{(24)} $ & $\left(0,-1,0,0,0,0,-{\frac{1}{{\sqrt{2}}}}\right)$\\ \hline
$G_{\mu 6}$ & $G_{\mu 6}$ & $W^{(25)} $ & $\left(0,0,-1,0,0,0,-{\frac{1}{{\sqrt{2}}}}\right)$\\ \hline
$G_{\mu 7}$ & $G_{\mu 7}$ & $W^{(26)} $ & $\left(0,0,0,-1,0,0,-{\frac{1}{{\sqrt{2}}}}\right)$\\ \hline
$G_{\mu 8}$ & $G_{\mu 8}$ & $W^{(27)} $ & $\left(0,0,0,0,-1,0,-{\frac{1}{{\sqrt{2}}}}\right)$\\ \hline
$B_{\mu 9}$ & $B_{\mu 9}$ & $W^{(28)} $ & $\left(0,0,0,0,0,1,-{\frac{1}{{\sqrt{2}}}}\right)$\\ \hline
$A_{\mu 456789}$ & $A_{\mu 45678}$ & $W^{(29)} $ & $-\frac{1}{2}\left(-1,-1,-1,-1,-1,
{\pm} 1\right) $\\ \hline
$A_{\mu 49}$ & $A_{\mu 4}$ & $W^{(30)} $ & $-\frac{1}{2}\left(-1,1,1,1,1,{\pm} 1,0 \right) $\\ \hline
$A_{\mu 59}$ & $A_{\mu 5}$ & $W^{(31)} $ & $-\frac{1}{2}\left(1,-1,1,1,1,{\pm} 1,0 \right) $\\ \hline
$A_{\mu 69}$ & $A_{\mu 6}$ & $W^{(32)} $ & $-\frac{1}{2}\left(1,1,-1,1,1,{\pm} 1,0 \right) $\\ \hline
$A_{\mu 79}$ & $A_{\mu 7}$ & $W^{(33)} $ & $-\frac{1}{2}\left( 1,1,1,-1,1,{\pm} 1,0\right) $\\ \hline
$A_{\mu 89}$ & $A_{\mu 8}$ & $W^{(34)} $ & $-\frac{1}{2}\left(1,1,1,1,-1,{\pm} 1,0 \right) $\\ \hline
$A_{\mu 45}$ & $A_{\mu 459}$ & $W^{(35)} $ & $-\frac{1}{2}\left(-1,-1,1,1,1,{\mp} 1,0 \right) $\\ \hline
$A_{\mu 46}$ & $A_{\mu 469}$ & $W^{(36)} $ & $-\frac{1}{2}\left(-1,1,-1,1,1,{\mp} 1,0 \right) $\\ \hline
$A_{\mu 47}$ & $A_{\mu 479}$ & $W^{(37)} $ & $-\frac{1}{2}\left( -1,1,1,-1,1,{\mp} 1,0 \right) $\\ \hline
$A_{\mu 48}$ & $A_{\mu 489}$ & $W^{(38)} $ & $-\frac{1}{2}\left(-1,1,1,1,-1,{\mp} 1,0 \right) $\\ \hline
$A_{\mu 58}$ & $A_{\mu 589}$ & $W^{(39)} $ & $-\frac{1}{2}\left(1,-1,1,1,-1,{\mp} 1,0 \right) $\\ \hline
$A_{\mu 68}$ & $A_{\mu 689}$ & $W^{(40)} $ & $-\frac{1}{2}\left(1,1,-1,1,-1,{\mp} 1,0 \right) $\\ \hline
$A_{\mu 78}$ & $A_{\mu 789}$ & $W^{(41)} $ & $-\frac{1}{2}\left( 1,1,1,-1,-1,{\mp} 1,0\right) $\\ \hline
$A_{\mu 57}$ & $A_{\mu 579}$ & $W^{(42)} $ & $-\frac{1}{2}\left(1,-1,1,-1,1,{\mp} 1,0 \right) $\\ \hline
$A_{\mu 56}$ & $A_{\mu 569}$ & $W^{(43)} $ & $-\frac{1}{2}\left(1,-1,-1,1,1,{\mp} 1,0 \right) $\\ \hline
$A_{\mu 67}$ & $A_{\mu 679}$ & $W^{(44)} $ & $-\frac{1}{2}\left(1,1,-1,-1,1,{\mp} 1,0 \right) $\\ \hline
$B_{\mu 4}$ & $B_{\mu 4}$ & $W^{(45)} $ & $\left(1,0,0,0,0,0,-{\frac{1}{{\sqrt{2}}}}\right)$\\ \hline
$B_{\mu 5}$ & $B_{\mu 5}$ & $W^{(46)} $ & $\left(0,1,0,0,0,0,-{\frac{1}{{\sqrt{2}}}}\right)$\\ \hline
$B_{\mu 6}$ & $B_{\mu 6}$ & $W^{(47)} $ & $\left(0,0,1,0,0,0,-{\frac{1}{{\sqrt{2}}}}\right)$\\ \hline
$B_{\mu 7}$ & $B_{\mu 7}$ & $W^{(48)} $ & $\left(0,0,0,1,0,0,-{\frac{1}{{\sqrt{2}}}}\right)$\\ \hline
$B_{\mu 8}$ & $B_{\mu 8}$ & $W^{(49)} $ & $\left(0,0,0,0,1,0,-{\frac{1}{{\sqrt{2}}}}\right)$\\ \hline
$G_{\mu 9}$ & $G_{\mu 9}$ & $W^{(50)} $ & $\left(0,0,0,0,0,-1,-{\frac{1}{{\sqrt{2}}}}\right)$\\ \hline
$G_{\mu 4}$ & $G_{\mu 4}$ & $W^{(51)} $ & $\left(1,0,0,0,0,0,{\frac{1}{{\sqrt{2}}}}\right)$\\ \hline
$G_{\mu 5}$ & $G_{\mu 5}$ & $W^{(52)} $ & $\left(0,1,0,0,0,0,{\frac{1}{{\sqrt{2}}}}\right)$\\ \hline
$G_{\mu 6}$ & $G_{\mu 6}$ & $W^{(53)} $ & $\left(0,0,1,0,0,0,{\frac{1}{{\sqrt{2}}}}\right)$\\ \hline
$G_{\mu 7}$ & $G_{\mu 7}$ & $W^{(54)} $ & $\left(0,0,0,1,0,0,{\frac{1}{{\sqrt{2}}}}\right)$\\ \hline
$G_{\mu 8}$ & $G_{\mu 8}$ & $W^{(55)} $ & $\left(0,0,0,0,1,0,{\frac{1}{{\sqrt{2}}}}\right)$\\ \hline
$G_{\mu 9}$ & $B_{\mu 9}$ & $W^{(56)} $ & $\left(0,0,0,0,0,-1,{\frac{1}{{\sqrt{2}}}}\right)$\\\hline
\end{tabular}}
\end{center}
\caption{\small Correspondence between the weights ${W^{(\lambda)}}^{\mp}$ of the ${\bf 56}$ of ${\cal E}_{7(7)}^{\pm}$ and the vectors deriving from the dimensional reduction of 
type IIA and type IIB fields.}
\label{solvec1}
\end{table}

\end{document}